\documentclass[pdflatex,12pt]{elsarticle}
\usepackage{amsmath}
\usepackage{amssymb}
\usepackage{amsfonts}
\usepackage{graphics}
\usepackage{graphicx}
\usepackage{amscd}
\usepackage{amsfonts}
\usepackage{epsf}
\usepackage{epsfig}
\usepackage{color}
\usepackage{feynmf}
\usepackage{wrapfig}

\usepackage{enumerate}
 \usepackage[utf8]{inputenc} 
 \usepackage{ascmac}%proof we added

\biboptions{numbers,sort&compress}
\usepackage{etoolbox}
\makeatletter
\patchcmd{\ps@pprintTitle}% <cmd>
  {\let\@oddhead\@empty}% <search>
  {\def\@oddhead{\mbox{}\hfill \footnotesize{RIKEN-iTHEMS-Report-23}}}% <replace>
  {}{}% <success><failure>
\makeatother
\newcommand{\n}{\nonumber}
\newcommand{\la}{\langle}
\newcommand{\ra}{\rangle}
\newcommand{\eref}[1]{(\ref{#1})}

\allowdisplaybreaks[1]
\begin{document}

\title{Perturbative analysis of the Wess-Zumino flow} 

\author{Daisuke Kadoh}
\address{Faculty of Sciences and Engineering, Doshisha University, Kyoto 610-0394, Japan}
\ead{dkadoh@mail.doshisha.ac.jp}

\author{Kengo Kikuchi}
\address{RIKEN iTHEMS, Wako, Saitama 351-0198, Japan}
\ead{kengo@yukawa.kyoto-u.ac.jp}

\author{Naoya Ukita}
\address{Center for Computational Sciences, University of Tsukuba, Tsukuba, Ibaraki 305-8577, Japan}
\ead{ukita@ccs.tsukuba.ac.jp}

\date{\today}

\begin{abstract}

We investigate an interacting supersymmetric gradient flow in the Wess-Zumino model. 
Thanks to the nonrenormalization theorem and an appropriate initial condition,  we find that 
any correlator of flowed fields is ultraviolet finite. 
This is shown at all orders of the perturbation theory using the power counting theorem for
one-particle irreducible supergraphs.
Since the model does not have the gauge symmetry,
 the mechanism of realizing the ultraviolet finiteness is 
quite different from that of the Yang-Mills flow, and  
this could provide further understanding of the gradient flow approach.

\end{abstract}

\maketitle

\newpage

\section{Introduction}
\label{sec:Intoroduction}

The gradient flow has achieved great success in lattice field theory\cite{Luscher:2010iy, Luscher:2011bx}, 
and there are many applications, such 
as nonperturbative renormalization group\cite{Yamamura:2015kva, Makino:2018rys, Abe:2018zdc, Carosso:2018bmz, Sonoda:2019ibh, Sonoda:2020vut, Miyakawa:2021hcx, Miyakawa:2021wus, Miyakawa:2022qbz, Sonoda:2022fmk, Hasenfratz:2022wll, Abe:2022smm}, 
a holographic description of field theory\cite{Aoki:2015dla, Aoki:2016ohw, Aoki:2017uce, Aoki:2017bru, Aoki:2018dmc, Aoki:2019bfb, Aoki:2022lye},
O($N$) nonlinear sigma model and large $N$ expansion\cite{Makino:2014cxa, Aoki:2014dxa, Makino:2014sta, Aoki:2016env}, supersymmetric theory\cite{Nakazawa:2003zf, Nakazawa:2003tz, Kikuchi:2014rla, Aoki:2017iwi, Kadoh:2018qwg, Kadoh:2019glu, Kadoh:2019flv, Bergner:2019dim, Hieda:2017sqq, Kasai:2018koz},
and phenomenological physics to obtain the bounce solution or 
sphaleron fields configuration\cite{Chigusa:2019wxb, Sato:2019axv, Hamada:2020rnp, Ho:2019ads}.
Further studies of the gradient flows 
could provide a deep understanding of field theories\cite{Fujikawa:2016qis,Morikawa:2018fek}.

In the Yang-Mills flow, any correlator of the flowed field is 
ultraviolet(UV) finite
at positive flow time
if the four-dimensional Yang-Mills 
theory is properly renormalized.
In the case of QCD, with an extra field strength renormalization 
for the flowed quarks, a similar property is obtained\cite{Luscher:2013cpa}.
This property is a key ingredient of the flow approach,    
and physical quantities that are difficult to define exactly on the lattice
can be studied by lattice simulations with the flows\cite{Suzuki:2013gza, Makino:2014taa, Asakawa:2013laa, Taniguchi:2016ofw, Kitazawa:2017qab, Yanagihara:2018qqg, Harlander:2018zpi, Iritani:2018idk}.

Such a UV finiteness of gradient flow, however, does not hold
for scalar field theory in general\cite{Capponi:2015ucc}. 
The interacting flow 
has nonremovable divergences, and 
the extra field strength renormalization remains 
even for the massless free flow.\footnote{
The flow equation is given only from the gradient of the massless free part of the action, 
while the scalar field theory at $t=0$ still has interaction terms. 
The initial condition is given by  a bare scalar field. }
This seems to suggest that 
the gauge symmetry or other symmetries are necessary in realizing 
the UV finiteness of the interacting gradient flow.

Supersymmetric gradient flow is another possibility of realizing the UV finiteness. 
The supersymmetric flows are constructed 
for the super-Yang-Mills in Refs.~\cite{Kikuchi:2014rla, Kadoh:2018qwg} and for the super-QCD in Ref.~\cite{Kadoh:2019flv}. 
In Ref.~\cite{Kadoh:2019glu}, we also constructed 
a supersymmetric flow in the 
Wess-Zumino model, which is referred to as Wess-Zumino flow in this paper.
The Wess-Zumino flow is the simplest supersymmetric extension of the gradient flow and gives a good testing ground 
in investigating the influence of supersymmetry on the flow approach.

%%%%%%%%%%%%%%%%

In this paper, 
we show that any correlation function of chiral superfields obtained from the Wess-Zumino flow  is 
UV finite at positive flow time in all orders of the perturbation theory.
Since the model does not have the gauge symmetry, the mechanism of realizing the 
UV finiteness is quite different from that of the Yang-Mills flow.
As we will see later, it is a direct consequence of the supersymmetry, in particular, the nonrenormalization 
theorem of the Wess-Zumino model.

To show this, we first introduce a method of defining a Wess-Zumino flow 
with renormalization-invariant couplings. 
We also give a renormalization-invariant initial condition.
These renormalization invariances are immediately shown 
from the nonrenormalization theorem.
The perturbation calculation of the Wess-Zumino flow is carried out using
an iterative expansion of the flow equation and the ordinary perturbation theory 
for the boundary Wess-Zumino model. 
Since the initial condition depends on the coupling constant, 
the order of the perturbative expansion is given by a fractional power $g^{2/3}$. 
The super-Feynman rule for one-particle irreducible (1PI) supergraphs is then derived. 
Using the power counting theorem based on the super-Feynman rule, 
the UV finiteness of the Wess-Zumino flow is established. 

The rest of this paper is arranged as follows: 
In Sec.~\ref{phi-four}, we consider the gradient flow of the $\phi^4$ scalar field theory. 
In Sec.~\ref{sec:WZ_review}, we review a perturbation theory in the Wess-Zumino model as a supersymmetric extension of $\phi^4$ scalar field theory.
In Sec.~\ref{sec:WZ_flow}, we construct the Wess-Zumino flow with renormalization-invariant couplings according to Ref.~\cite{Kadoh:2019glu} with some modifications.  
With the super-Feynman rule for the correlation function derived from the iterative expansion of the flow equation,
we show that the Wess-Zumino flow has UV finiteness using the power counting theorem.  Section \ref{sec:summary} is devoted to summarizing results.

%%%%%%%%%%%%%%%%

\section{The case of $\phi^4$ theory}
\label{phi-four}

Let $t \ge 0$ be a flow time and $\varphi(t,x)$ be a $t$-dependent field. 
We consider a gradient flow equation of Euclidean $\phi^4 $ theory as 
\begin{align}
\frac{\partial\varphi(t,x)}{\partial t} = (\Box -m^2)\varphi (t,x) - \lambda \varphi^3(t,x)
\label{scalar_flow}
\end{align}
with an initial condition,  
\begin{align}
\varphi(t=0,x)=\phi(x)\label{initial2},
\end{align}
where $\Box = \partial_\mu \partial_\mu$. 
As the name suggests, 
the rhs of Eq.~(\ref{scalar_flow})
 is $-\delta S/\delta \phi|_{\phi \rightarrow \varphi}$ 
where 
\begin{align}
S=\int d^4 x \,  \left\{ \frac{1}{2}(\partial_\mu  \phi)^2
+ \frac{m^2}{2}\phi^2+\frac{\lambda}{4}\phi^4
\right\}(x)
\label{scalar_action}
\end{align}
with a bare mass $m$ and a bare coupling constant $\lambda$. 
In this setup, the scalar theory (\ref{scalar_action}) is put on the boundary ($t=0$).

In the Yang-Mills flow, it is shown that correlation functions at positive flow time are
UV finite under the initial condition $B_\mu(t=0,x)=A_\mu(x)$ where
$A_\mu(x)$ is a bare field irrelevant to a renormalization scheme. 
This property plays a crucial role in matching two different schemes that are used 
for calculating nontrivial renormalizations for operators\cite{Luscher:2010iy, Luscher:2013cpa, Suzuki:2013gza, Makino:2014taa}. 
In this paper, we also employ an initial condition given by bare fields for the Wess-Zumino flow 
in later sections, such as Eq.~\eqref{initial2} for scalar theory.
%%%

The formal solution of Eq.~(\ref{scalar_flow}) can be obtained from an 
iterative approximation of the flow equation. 
This is regarded as a perturbative expansion in terms of $\lambda$.
The flowed field $\varphi(t,x)$ is thus given by a treelike graph with the boundary field $\phi$ at the end points.  
The correlation function of the flowed field is then evaluated by the usual perturbation  
theory at the boundary\cite{Luscher:2010iy, Luscher:2011bx}.

%%%%%%%%%%%%
In the massless free flow where $\partial\varphi/\partial t = \Box\varphi$ 
and Eq.\eqref{scalar_action} gives the boundary theory,\footnote{ 
In this case, the action that defines the gradient flow is different from the boundary theory.}
any correlation function of $\varphi(t,x)$ 
is UV finite up to an extra wave function renormalization 
once the boundary theory is properly renormalized.
However, for massive or interacting flows ($m \neq 0$ or $\lambda\neq 0$),
such a property is not obtained\cite{Capponi:2015ucc}.

This conclusion is easily understood from the 4+1-dimensional theory that 
produces the same perturbative series discussed above.  
As in the case of  the Yang-Mills flow\cite{Luscher:2011bx},
the bulk action of the 4+1-dimensional theory is given 
by
\begin{align}
& S_{\mathrm{bulk}}=\int_0^{\infty}dt \int d^4 x 
L(t,x) \bigg\{  \partial_t\varphi   (t,x) - (\Box-m^2)\varphi (t,x) + \lambda \varphi^3 (t,x) \bigg\}
\label{bulk_action}
\end{align}
with a Lagrange multiplier field $L(t,x)$. The effect of the boundary field on the bulk field $\varphi(t,p)$ 
is suppressed by a damping factor ${\rm e}^{-t p^2}$. 
Therefore, at large flow times, correlation functions of the bulk fields are given by Feynman diagrams 
consisting only of flow lines and flow vertices. Any diagram of this kind resulting from the action \eqref{bulk_action} starts from $L$ and ends at $\varphi$, and is expressed as a directed graph without loops. 
Since there are no divergences, 
bulk counterterms are absent for the action \eqref{bulk_action}. 
However, $m$ and $\lambda$ are the bare parameters of the boundary theory 
and contain divergences determined in the theory. 
Therefore, unnecessary ``bulk counterterms'' arise from the renormalization of $m$ and $ \lambda$, and this $d+1$-dimensional theory is nonrenormalizable.
\footnote{In the massless free flow,
there are no ``bulk counterterms'', and 
any UV divergence of flowed field correlators appears  only
in loop integrals at the boundary. 
If we took $\varphi(t=0,x)=\phi_{R}(x)$ instead of Eq.~\eref{initial2}, 
any correlation function is UV finite.}

Achieving UV finiteness 
in the massive or interacting flow requires the absence of the bulk counterterms. 
In other words, the flow equation should be given by renormalization-invariant couplings.
We consider a supersymmetric $\phi^4$ theory in the next section 
because further constraints on the renormalization are needed to define such a renormalization-invariant flow equation. 

%%%%%%%%%%%%%%%%%%%%%%%%%%%%%%%%%%%%%%%%%%
%
%
%            Section 3
%
%
%%%%%%%%%%%%%%%%%%%%%%%%%%%%%%%%%%%%%%%%%%

\section{Review of the Wess-Zumino model}
\label{sec:WZ_review}

We work in Euclidean space with the notation of Refs.~\cite{Kadoh:2018qwg, Kadoh:2019glu},
which is derived from Ref.~\cite{Wess:1992cp} by a Wick rotation.
See \ref{notation} for details of the notation.

\subsection{The Wess-Zumino model}
The Wess-Zumino model is a supersymmetric extension of $\phi^4$ theory, which is 
given by a scalar field $A(x)$, a Weyl spinor $\psi(x)$, and an auxiliary field $F(x)$. 
In the superfield formalism,
 a chiral superfield 
$\Phi(x,\theta,\bar\theta)$ contains the field contents as
\begin{align}
\Phi(y,\theta) \equiv A(y)+\sqrt{2}\theta\psi(y) + i\theta\theta F(y),
\end{align}
where $y_\mu=x_\mu+ i\theta \sigma_\mu \bar\theta$. 

In Minkowski space, an antichiral superfield $\Phi^\dag$ is defined by the Hermitian conjugate of $\Phi$. 
However, in Euclidean space, such a definition is incompatible with the Wick rotation. 
In fact, $\bar \psi$ is not a Hermitian conjugate of $\psi$ but a different Weyl spinor. 
We define an antichiral superfield $\bar \Phi$ that is a Euclidean counterpart of $\Phi^\dag$ as
\begin{align}
\bar \Phi(\bar y,\bar \theta) \equiv A^\ast(\bar y)+\sqrt{2}\bar \theta\bar\psi(\bar y) + i\bar\theta\bar\theta F^\ast (\bar y),
\end{align}
where  $\bar y_\mu=x_\mu- i\theta \sigma_\mu \bar\theta$.

In Euclidean space, the chiral and antichiral superfields $\Phi$ and $\bar\Phi$ also satisfy 
$\bar D_{\dot\alpha} \Phi=D_{\alpha} \bar\Phi=0$. The  supersymmetry transformation of a superfield 
${\cal F}(x,\theta,\bar\theta)$ is defined as
\begin{align}
\delta_\xi {\cal F}(x,\theta,\bar\theta) = (\xi Q + \bar\xi \bar Q) {\cal F}(x,\theta,\bar\theta),
\label{super_transf}
\end{align}
where the supercovariant derivatives $D, \bar D$ and supercharges $Q, \bar Q$ are defined in 
\ref{notation}. 
Supersymmetry transformations of component fields are derived from \eqref{super_transf}.

The Wess-Zumino model  is then defined by 
\begin{align}
&& S =-\int d^8 z \, \bar\Phi(z)
 \Phi(z)   
-\int d^4x d^2 \theta W(\Phi(z)) - \int d^4x d^2 \bar\theta W(\bar\Phi(z)),
\label{wz_action}
\end{align}
where 
 \begin{align}
W(\Phi) \equiv  \frac{m}{2}\Phi^2+\frac{g}{3} \Phi^3
\end{align}
for bare coupling constants $m\ge 0$ and $g >0$.
To simplify the notation, we used $z=(x_\mu,\theta_\alpha, \bar\theta_{\dot\alpha})$ and 
$d^8 z \equiv d^4 x d^2\theta d^2\bar\theta$. 
The action is invariant under the supersymmetry transformation (\ref{super_transf}). 

Renormalized superfield $\Phi_R$ and renormalized coupling constants $m_R, g_R$ satisfy 
\begin{align}
\Phi_R = Z^{-\frac{1}{2}} \Phi, \quad \ \ \bar\Phi_R = Z^{-\frac{1}{2}} \bar\Phi,
\label{ren_super_field}
\end{align}
and
\begin{align}
\delta_m = mZ- m_R, \quad  \delta_g=g Z^{\frac{3}{2}} -g_R.
\end{align}
The nonrenormalization theorem of the Wess-Zumino model 
tells us that the F-terms are not renormalized, that is, $\delta_m=\delta_g=0$ 
\cite{Wess:1973kz, Iliopoulos:1974zv, Grisaru:1979wc, Seiberg:1993vc}. 
Therefore, we have
\begin{align}
m_R = mZ, \qquad  g_R=g Z^{\frac{3}{2}}.
\label{non-renormalization}
\end{align}
It turns out that a normalized mass given by
\begin{align}
M \equiv mg^{-\frac{2}{3}}
\label{inv_mass}
\end{align}
is invariant under the renormalization.

\subsection{Perturbation theory}\label{PT}

The perturbation theory can be given 
in the superfield formalism\cite{Grisaru:1979wc}. 
We derive a super-Feynman rule for 1PI supergraphs of 
the Wess-Zumino model in Euclidean space. 
Equation~\eqref{non-renormalization} is formally confirmed by the power counting theorem derived from the super-Feynman rule.

We first introduce external chiral and antichiral superfields
$J$ and $\bar{J}$ satisfying $\bar D_{\dot\alpha} J=D_\alpha\bar{J}=0$ and consider
\begin{align}
Z[J,\bar{J}]=\int D\Phi D\bar{\Phi} \, e^{-S_0 -S_{int}-S_{src} },
\label{ZJ}
\end{align}
where
\begin{align}
S_{src} = -\int d^4 x d^2\theta \, J(z) \Phi(z) - \int d^4 x d^2\bar\theta \, \bar J(z) \bar \Phi(z).
\end{align}
The superfield Green's function $G(z_1,z_2,\cdots,z_m; z^\prime_{1},z^\prime_{2},\cdots, z^\prime_{n})$ is obtained by
\begin{align}
& G(z_1,\cdots,z_m; z^\prime_{1},\cdots, z^\prime_n)  \nonumber \\
& \hspace{1cm} = \frac{1}{Z|_{J=\bar J=0}}
\left. \frac{\delta^m}{\delta J (z_1)\cdots \delta J (z_m)} 
\frac{\delta^n}{\delta \bar J (z^\prime_{1}) \cdots \delta \bar J (z^\prime_{n})} \, 
 Z[J,\bar{J}] \right|_{J=\bar J=0},
 \label{green_function}
\end{align}
where 
\begin{align}
\frac{\delta J(z_1)}{\delta J(z_2)}=&-\frac{\bar D_1^2}{4}\delta^8(z_1-z_2),  \\
\frac{\delta \bar{J}(z_1)}{\delta \bar{J}(z_2)}=&-\frac{D_1^2}{4}\delta^8 (z_1-z_2),
\end{align}
and the other functional derivatives are zero, 
where $D_1$ and $\bar D_1$ are defined for $z_1$.

Let $S_0$ and $S_{int}$ be  the free and interaction parts of the action, respectively. 
The free field action $S_0$ can be written in the full superspace as 
\begin{align}
S_0=-&\int d^8z \left\{ \bar\Phi \Phi + \frac{m}{2}\Phi \left(-\frac{D^2}{4\Box}\right) \Phi
+\frac{m}{2}\bar\Phi \left(-\frac{\bar{D}^2}{4\Box}\right)\bar\Phi \right\}(z).
\end{align}
Similarly, we have
\begin{align}
S_{int}=-\frac{g}{3} \int d^8z \left\{ \Phi^2 \left(-\frac{D^2}{4\Box} \right) \Phi + \bar \Phi^2 \left(-\frac{\bar{D}^2}{4\Box}\right)
\bar\Phi \right\}(z),
\label{int_wz}
 \end{align}
and 
\begin{align}
S_{src} = -\int d^8z \left\{ J\left(-\frac{D^2}{4\Box}\right) \Phi
+\bar J \left(-\frac{\bar{D}^2}{4\Box}\right)\bar\Phi \right\}(z).
\end{align}
These are easily derived using Eqs.~\eref{identity_Box1} and \eref{identity_Box2}. 

A short calculation tells us that $Z_0[J,\bar J] \equiv Z[J,\bar J]|_{g=0}$ is written as 
\begin{align}
Z_0[J,\bar{J}] 
=&\exp\left\{\frac{1}{2}\int d^8z d^8z'  \left(
J(z), \bar{J}(z)
\right) \Delta_{GRS}(z,z')\left(
\begin{array}{cc}
J(z^\prime)\\
\bar{J}(z^\prime)
\end{array}
\right) \right\},
\label{GRS_Z}
\end{align}
where 
 \begin{align}
\Delta_{GRS}(z,z')=&\frac{1}{-\Box+m^2}\left(
\begin{array}{cc}
\frac{m D^2}{4\Box}&1\\
1&\frac{m \bar{D}^2}{4\Box}
\end{array}
\right)\delta^8(z-z').
\end{align}
The propagator $\Delta_{GRS}$ is called
the Grisaru-Rocek-Siegel (GRS) propagator introduced in \cite{Grisaru:1979wc}.

Two-point functions are thus obtained as
\begin{align}
&\langle\Phi(z_1)\bar\Phi(z_2) \rangle_0 =\frac{1}{16} \frac{\bar D^2_1 D_1^2}{-\Box_1+m^2} \delta^8(z_1-z_2),\n \\
&\langle\Phi(z_1)\Phi(z_2) \rangle_0 = \frac{m}{4} \frac{\bar D^2_1}{-\Box_1+m^2} \delta^8(z_1-z_2),
\label{Phi_correlation_function}
\\
&\langle\bar\Phi(z_1)\bar\Phi(z_2) \rangle_0 =\frac{m}{4} \frac{D^2_1}{-\Box_1+m^2} \delta^8(z_1-z_2), \n
\end{align}
where $\langle \cdots  \rangle_0$ is the expectation value in the free theory. 
The Green's function \eqref{green_function}  is obtained from
\begin{align}
Z[J,\bar{J}]= \exp\left\{-S_{int}\left[\frac{\delta}{\delta J}, \frac{\delta}{\delta \bar{J}}     \right]  \right\}
Z_0[J,\bar{J}],
\end{align}
by evaluating the functional derivatives $\delta/\delta J$ and $\delta/\delta \bar J$. 
In perturbation theory, we need to evaluate extra derivatives that arise from the Taylor expansion of 
$ \exp\left\{-S_{int}\left[\frac{\delta}{\delta J}, \frac{\delta}{\delta \bar{J}}     \right]  \right\}$.

The perturbative calculation of Green's functions contains a term like
 \begin{align}
&-S_{int}\left[\frac{\delta}{\delta J},0 \right] J(z_1)J(z_2)J(z_3)\n\\
& =  \frac{g}{3} \int d^8 z_4 \left\{-\frac{D_4^2}{4\Box_4} \left(\frac {\delta}{\delta J(z_4)}\right)\right\}\left(\frac {\delta}{\delta J(z_4)}\right)^2J(z_1)J(z_2)J(z_3)\n\\
  &= 2g \int d^8 z_4 \delta^8(z_1-z_4) \left(-\frac{\bar D_2^2}{4} \right)\delta^8(z_2-z_4) 
  \left(-\frac{\bar D_3^2}{4} \right)\delta^8 (z_3-z_4),
\label{sample}
\end{align}
where $J(z_i)$ attaches to antichiral superfields via Eq.~\eref{GRS_Z}. 
We used \eref{identity_Box2} to show the second equality. 

The effective action is made of 1PI supergraphs
 that are calculated from 1PI Green's functions amputating propagators of external lines.
Each vertex of 1PI diagrams has two or three internal lines. 
For a vertex with no external lines, two of the three internal lines have $\frac{\bar D^2}{4}$ as suggested 
from the last line of Eq. \eqref{sample}.
 Whereas, for a vertex with two internal lines and one external line, 
  one of the two internal lines has $\frac{\bar D^2}{4}$ 
  because the external lines are associated 
  with $\delta/\delta J$ without $\frac{D^2}{4\Box}$ in the second line of \eqref{sample}.

The super-Feynman rules for 1PI supergraphs are given in the momentum space as follows: 
 \begin{enumerate}[ \ \  (a)]
\item Use the propagators $\tilde\Delta_{GRS}$ for $\Phi\Phi, \Phi\bar\Phi, \bar\Phi\bar\Phi$, which are given by 
\begin{align}
\tilde\Delta_{GRS}(p;\theta_1,\bar\theta_1,\theta_2,\bar\theta_2)  \n \\
& \hspace{-3cm} =  
\frac{1}{p^2+m^2}\left(
\begin{array}{cc}
-\frac{m D_1^2}{4p^2}&1\\
1&-\frac{m \bar{D}_1^2}{4p^2}
\end{array}
\right)\delta^2(\theta_1-\theta_2)\delta^2(\bar\theta_1-\bar\theta_2).
\label{GRS_mom}
\end{align}
\item Write a factor $2g$ and $\int d^2 \theta d^2\bar\theta$ at each vertex. 
For a vertex with $n$ internal lines ($n=2,3$), 
put a factor of $-\bar{D}^2/4$  $(-D^2/4)$ at $n-1$ lines of the $n$ chiral (antichiral) lines. 
\item Impose the momentum conservation at each vertex and integrate over undetermined loop momenta.
\item Compute the usual combinatoric factors. 
\end{enumerate}
These rules are given in Euclidean space. See also Ref.~\cite{Wess:1992cp}  for the rule in Minkowski space.

We can calculate the superficial degrees of divergence for 1PI supergraphs using the super-Feynman rule. 
Consider a 1PI supergraph with $L$ loops, $V$ vertices, $E$ external lines
and $P$ propagators of which $C$ are $\Phi\Phi$ or $\bar\Phi\bar\Phi$ massive propagators.
We count $D^2,\bar D^2$ as $p$ because $\bar D^2 D^2 \sim p^2$ for chiral superfields. 
Each loop integral has  $d^4p$. 
The GRS propagator provides $1/p^2$ with an additional factor $1/p$ for $\Phi\Phi$ or $\bar\Phi\bar\Phi$ propagators. 
The internal lines have $2V-E$ factors of  $D^2$ or $\bar D^2$. 
In each loop integral, we can use an identity $\delta_{12} D^2 \bar D^2 \delta_{12}=16 \delta_{12}$ 
to remove a $D^2\bar D^2 \sim p^2$. 
The superficial degrees of divergence for the graph is given by
\begin{align}
d = 4L -2P - C + 2V - E - 2L.
\end{align}
Using  $V-P+L=1$, we find 
\begin{align}
d = 2-E-C. 
\label{sdiv}
\end{align}
For $E=2$, $d$ can be zero (the logarithmic divergence). 
If two external lines have the same chirality, $d<0$ 
because at least one $\Phi\Phi$ or $\bar\Phi\bar\Phi$ propagator is needed. 
We have $d<0$ for $E \ge 3$.
Thus we find that the wave function renormalization exists 
but the effective action does not have any divergent correction to $m\Phi^2$ and $g \Phi^3$.

%%%%%%%%%%%%%%%%%%%%%%%%%%%%%%%%%%%%%%%%%%
%
%
%            Section 4
%
%
%%%%%%%%%%%%%%%%%%%%%%%%%%%%%%%%%%%%%%%%%%

\section{The Wess-Zumino flow}\label{sec:WZ_flow}
We consider a supersymmetric gradient flow in the Wess-Zumino model.
It can be shown that any correlation function of the flowed fields is UV finite 
thanks to the nonrenormalization theorem under an appropriate initial condition.

%%%%%%%%%%%%%%%%%%%%%%%%%%%%%%%%%%%%%%%%%%
%
%
%            Section 4.1 
%
%
%%%%%%%%%%%%%%%%%%%%%%%%%%%%%%%%%%%%%%%%%%

\subsection{The Wess-Zumino flow with renormalization-invariant couplings}

In Ref.~\cite{Kadoh:2019glu}, we defined a supersymmetric flow equation 
using the gradient of the action \eref{wz_action}. 
However, the bulk counterterms exist in this case,
because the bare coupling constants $m, g$ included in the flow
receive the renormalizations determined at $t=0$. 
See Ref.~\cite{Capponi:2015ucc} for relevant arguments. 
Therefore, the flow theory with bare $m$ and $g$ is ill defined at the quantum level.

In order to solve this issue, we introduce renormalization-invariant couplings into the flow equation.
We consider the following rescaling of coordinates and field variables:
\begin{align}
 \begin{split}
x^\prime_\mu \equiv g^{\frac{2}{3}} x_\mu, \quad
\theta^\prime \equiv g^{\frac{1}{3}}\theta, \quad
\bar\theta^\prime \equiv g^{\frac{1}{3}}\bar\theta 
 \end{split}
\end{align}
and 
\begin{align}
&A^\prime (x^\prime)\equiv g^{\frac{1}{3}}A(x),  \nonumber \\
&\psi^\prime(x^\prime) \equiv \psi(x),  \\
\label{rescaling}
& F^\prime (x^\prime) \equiv g^{-\frac{1}{3}} F(x).  \nonumber
\end{align}
Replacing every variable of the superfields by the corresponding rescaled variable, 
we have
\begin{align}
 \Xi (x^\prime, \theta^\prime, \bar\theta^\prime) \equiv  g^{\frac{1}{3}} \Phi (x,\theta,\bar\theta), 
\end{align}
where 
\begin{align}
\Xi
 (y^\prime,\theta^\prime) =
 A'(y')+\sqrt{2}\theta'\psi'(y')+ i\theta'\theta' F'(y'), \label{prime}
\end{align}
and  $y^\prime_\mu \equiv x'_\mu+i\theta'\sigma_\mu\bar\theta' = g^{\frac{2}{3}}y$.
The differential operators satisfy $Q^\prime_\alpha =  g^{\frac{1}{3}}Q_\alpha$  and $D^\prime_\alpha =  g^{\frac{1}{3}}D_\alpha$. The superfield formalism is then kept unchanged because 
$\Xi$ is a chiral superfield satisfying $\bar D'_{\dot \alpha} \Xi =0$ and 
the supersymmetry transformation laws of $A^\prime, \psi^\prime, F^\prime$ are the same as those of 
 $A, \psi, F$.

Hereafter, we omit the prime symbols unless they are confusing. 
From a short calculation, one can show that
the Wess-Zumino action is rewritten in  $\Xi(x,\theta,\bar\theta)$ and $\bar\Xi(x,\theta,\bar\theta)$ as
\begin{align}
S =&-\frac{1}{g^2} \int d^4x d^2\theta d^2\bar\theta  \bar\Xi \Xi  
-   \frac{1}{g^2}  \int d^4x d^2\theta\left(\frac{1}{2} M \Xi^2+\frac{1}{3}\Xi^3 \right)\n\\
 & -   \frac{1}{g^2}  \int d^4x d^2\bar\theta\left(\frac{1}{2} M \bar\Xi^2+\frac{1}{3}\bar\Xi^3 \right).
\label{wz_action_prime}
\end{align}
We should note that $M$ is defined as 
Eq.\eqref{inv_mass}, which is invariant 
under the renormalization for (\ref{wz_action}) in the standard manner.

In terms of rescaled variables, we can
consider a supersymmetric gradient flow according to Ref.~\cite{Kadoh:2019glu} as   
\begin{align}
\partial_t \Psi (t,z)  =
g^2\frac{\bar D^2}{4}  \frac{\delta S}{\delta \Xi (z)} \bigg|_{\Xi(z) \rightarrow \Psi(t,z)}
\label{superflow_formal_0},
\end{align}
where $z=(x,\theta,\bar\theta)$.  
The $\bar D^2$ factor is needed to keep the superchiral condition for
$\Psi (t,z)$ because ${\delta S}/{\delta \Xi}$  is not chiral. The flow equation for $\bar\Psi$ is given by a replacement 
$(\Psi,\Xi, \bar D) \leftrightarrow (\bar\Psi, \bar\Xi, D)$  from Eq.~(\ref{superflow_formal_0}).
We thus have 
\begin{align}
& \partial_t \Psi   = \Box\Psi   - M \frac{\bar D^2}{4} \bar \Psi  - \frac{ \bar D^2}{4} \bar{\Psi}^2,  \label{superflow1} \\
& \partial_t \bar \Psi   = \Box \bar \Psi   - M \frac{D^2}{4} \Psi  - \frac{ D^2}{4} {\Psi}^2 \label{superflow2}.
\end{align}
The flow equation is given with couplings that are renormalization invariant for the
original Wess-Zumino action \eqref{wz_action} given by $(A, \psi, F)$. 

The initial condition for $\Psi(t,z)$ and  $\bar \Psi(t,z)$ is given in the next section. 
If a supersymmetry transformation of the flowed fields
is defined 
by extending (\ref{super_transf}) to the 4+1 dimensions as 
$\delta_\xi \Psi (t,z) = (\xi Q + \bar\xi \bar Q) \Psi(t,z)$, then the flow equations and the supersymmetry transformation are consistent because 
they satisfy $[\delta_\xi,\partial_t]=0$.

The superchiral condition  $\bar D_{\dot\alpha} \Psi = D_{\alpha} \bar\Psi =0$ allows us to expand
$\Psi$ and $\bar \Psi$ as
\begin{align}
& \Psi(t,y,\theta) =  \phi(t,y) + \sqrt{2}\theta {\cal\chi}(t,y) + i \theta\theta G(t,y), \\
%\\ 
& \bar \Psi(t,\bar y,\theta) = \bar \phi(t,\bar y) + \sqrt{2}\bar\theta\bar\chi(t,\bar y) + i \bar\theta\bar\theta \bar G(t,\bar y).
\end{align}
For the component fields, we have 
\begin{align}
\partial_t \phi & = \Box \phi +i M \bar G +\left( 2 i \bar \phi \bar G -\bar\chi \bar\chi \right), 
\label{flow_phi}
\\
\partial_t \bar \phi  &= \Box \bar \phi +i M G  + \left( 2 i \phi G -\chi  \chi \right), 
\label{flow_barphi} \\
\partial_t \chi  & = \Box\chi  +i\sigma_\mu\partial_\mu \left(M \bar\chi+2  \bar \phi  \bar\chi \right),  
\label{flow_chi}
\\
\partial_t \bar\chi & = \Box\bar\chi +i\bar\sigma_\mu\partial_\mu \left(M \chi + 2 \phi \chi \right),   
\label{flow_barchi}
\\
 \partial_t G &= \Box  G-i \Box\left( M \bar \phi +  \bar \phi^{2} \right),
\label{flow_G} \\
 \partial_t \bar{G} &= \Box \bar G -i \Box \left(M \phi  +  \phi^{2}  \right).
\label{flow_barG}
\end{align}
Since the reality condition is broken by the Wick rotation, 
the Hermitian conjugate relation is not kept for the flow equation. 
So $\bar \phi$ and $\bar G$ are independent complex fields that are not complex conjugates of $\phi$ and $G$. 
From the initial condition given in the next section,
the complex conjugate relation is kept only at the boundary such as $\bar \phi(t=0,x)=(\phi(t=0,x))^*$. 
Note that the flow equations for $\bar \phi,\bar\chi,\bar G$ are obtained from those of $\phi,\chi, G$ 
by a simple replacement as $\phi \leftrightarrow \bar \phi, \chi \leftrightarrow \bar\chi, G \leftrightarrow \bar G$ and $\sigma_\mu \leftrightarrow \bar\sigma_\mu$.

%%%%%%%%%%%%%%%%%%%%%%%%%%%%%%%%%%%%%%%%%%
%
%
%            Section 4.2 
%
%
%%%%%%%%%%%%%%%%%%%%%%%%%%%%%%%%%%%%%%%%%%

\subsection{The vector notation and an initial condition}

We introduce a vector notation of chiral superfields as 
\begin{align}
{\bf \Psi}(t,z) =
\left(
\begin{array}{c}
{\bf \Psi}_1(t,z) \\
{\bf \Psi}_2(t,z) 
\end{array}
\right)
\equiv 
\left(
\begin{array}{c}
\Psi(t,z) \\
\bar \Psi(t,z)
\end{array}
\right).
\label{vector_Psi}
\end{align}
The Wess-Zumino flow equations \eqref{superflow1} and \eqref{superflow2} 
can be expressed as 
\begin{align}
\partial_t {\bf \Psi} = (\Box +M {\bf \Gamma \Delta}) {\bf \Psi} + \bar{\bf \Delta} {\bf N},
\label{WZflow_vector}
\end{align}
where 
\begin{align}
&{\bf \Delta} \equiv 
\left(
\begin{array}{cc}
 -\frac{1}{4} D^2 &  0  \\
0   & -\frac{1}{4}\bar D^2
\end{array}
\right),
\label{Delta}
\\
&\bar  {\bf \Delta} \equiv 
\left(
\begin{array}{cc}
   -\frac{1}{4} \bar D^2 & 0\\
0 & -\frac{1}{4} D^2      
\end{array}
\right),
\label{DeltaBar}
\\
&{\bf \Gamma} \equiv 
\left(
\begin{array}{cc}
   0 & 1\\
   1 & 0      
\end{array}
\right),
\label{Gamma}
\end{align}
and 
the nonlinear part is characterized by 
\begin{align}
{\bf N}_i (t, z) =\frac{1}{2}g_{ijk} {\bf \Psi}_j(t,z) {\bf \Psi}_k(t,z),
\end{align}
with a coefficient $g_{ijk}$ defined as $g_{ijk}=2{\bf\Gamma}_{ij} {\bf\Gamma}_{ik}$.

We consider the following initial condition,\footnote{
For the component fields, we have 
$\phi|_{t=0}  =\alpha A, \chi|_{t=0}  = \alpha \psi$, and $G|_{t=0}  = \alpha F$,
where $\alpha=g^\frac{1}{3}$. 
} 
\begin{align}
{\bf \Psi}|_{t=0} = {\bf \Phi}_0,
\label{inital_condition_Psi}
\end{align}
where 
\begin{align}
{\bf \Phi}_0(z) \equiv
g^{\frac{1}{3}}
\left(
\begin{array}{c}
\Phi(z) \\
\bar \Phi(z)
\end{array}
\right)=
g_R^{\frac{1}{3}}
\left(
\begin{array}{c}
\Phi_R(z) \\
\bar \Phi_R(z)
\end{array}
\right).
\label{inital_condition_Phi}
\end{align}
The second equality of Eq.(\ref{inital_condition_Phi})
is a direct consequence of the nonrenormalization theorem. 
We may consider ${\bf \Psi}|_{t=0}= f(M) {\bf \Phi}_0$ instead of Eq.(\ref{inital_condition_Psi}) 
because the conclusion of this section does not change for any nonzero function $f(M)$. 
Hereafter, we take $f(M)=1$ for simplicity.

The operators introduced above satisfy
\begin{align}
&\bar  {\bf \Delta} {\bf  \Delta} \bar  {\bf \Delta } = \Box \bar  {\bf \Delta }, \\
& {\bf \Gamma} \bar  {\bf \Delta}   {\bf \Gamma} =  {\bf \Delta}, \\ 
& {\bf \Gamma}^2={\bf 1}, 
\end{align}
and 
\begin{align}
\bar  {\bf \Delta}{\bf \Delta }  {\bf \Psi}  = \Box  {\bf \Psi} .
\end{align}

%%%%%%%%%%%%%%%%%%%%%%%%%%%%%%%%%%%%%%%%%%
%
%
%            Section 4. 3
%
%
%%%%%%%%%%%%%%%%%%%%%%%%%%%%%%%%%%%%%%%%%%

\begin{figure}[]
\begin{center}
\includegraphics [width=100mm]
{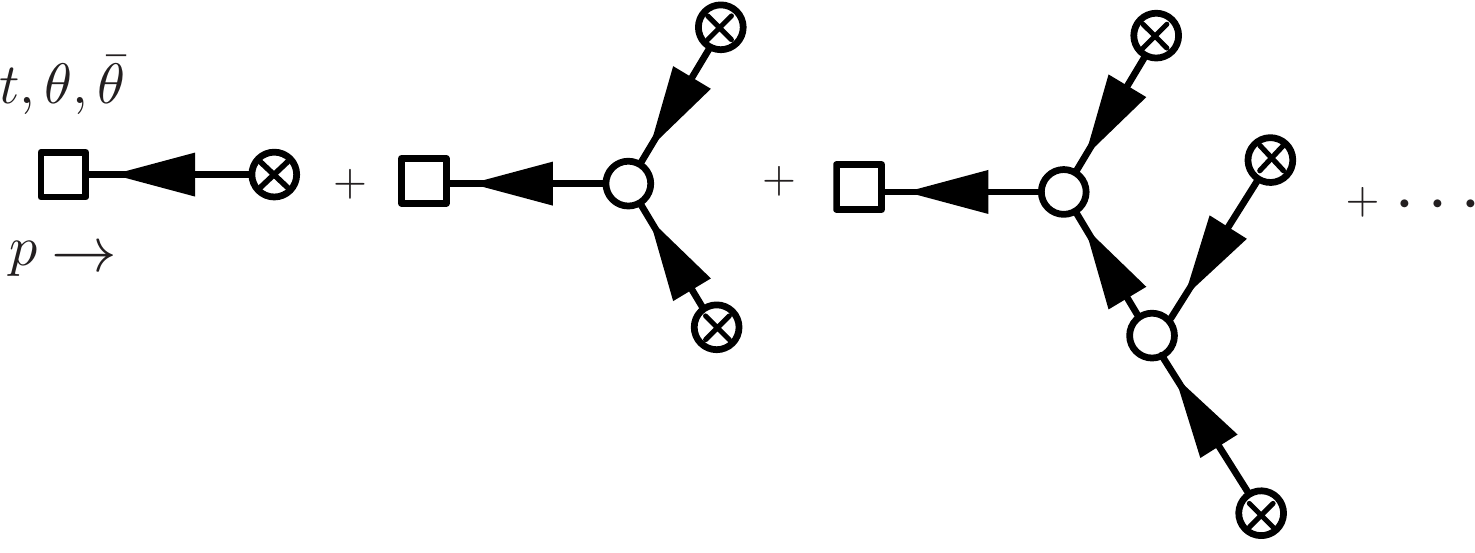}
\end{center}
\caption{Treelike graphs of the iterative solution ${\bf \Psi}(t,p,\theta,\bar\theta)$. }
\label{fig1}\end{figure}

\subsection{Iterative solution of the Wess-Zumino flow}
\label{iso}
The flowed field $\Psi(t,z)$ satisfying the Wess-Zumino flow equation can be expressed as an iterative expansion. 
To show this, we first introduce a heat kernel in the superspace $z=(x_\mu,\theta_{\alpha},\bar\theta_{\dot\alpha})$ as 
\begin{align}
{\bf K}_t (z) 
=\left(
\begin{array}{cc}
 C_t(x) &
 -\frac{\bar{D}^2}{4\sqrt{-\Box}} S_t(x)\\
 -\frac{D^2}{4\sqrt{-\Box}} S_t(x) & 
C_t(x)
\end{array}
\right) \times \delta^2(\theta)\delta^2(\bar\theta)
\end{align}
where
 \begin{align}
C_t(x)&\equiv \int \frac{d^4 p}{(2\pi)^4} e^{ipx-tp^2}\cos(tM \sqrt{p^2}),\\
S_t(x)&\equiv \int \frac{d^4 p}{(2\pi)^4} e^{ipx-tp^2}\sin(tM \sqrt{p^2}).
\end{align} 
The heat kernel satisfies
\begin{align}
\left(\partial_t -\Box - M{\bf \Gamma \Delta}\right) {\bf K}_t (z) =0
\end{align}
and 
\begin{align}
{\bf K}_0 (z) = \delta^8(z),
\end{align}
since $C_0(x) =\delta^4(x)$ and  $S_0(x)=0$.
The flow equation \eqref{WZflow_vector}
 can be solved formally as
\begin{align}
{\bf \Psi}(t,z) = \int d^8 z^\prime {\bf K}_t (z-z^\prime) {\bf \Phi}_0(z^\prime)
+ \int_0^t ds \int d^8 z^\prime \bar{\bf \Delta }{\bf K}_{t-s} (z-z^\prime) {\bf N}(s,z^\prime),
\label{formal_solution}
\end{align}
where $\bar{\bf \Delta}$ acts on $z$. 
Inserting the formal solution into ${\bf \Psi}$ of ${\bf N}$ on the rhs repeatedly 
yields an iterative approximation of the flow equation. 
The iterative approximation can be expressed as a treelike graph with ${\bf \Phi}_0$ 
at end points. 

In Fig.~\ref{fig1}, 
the iterative solution of the Wess-Zumino flow equation \eqref{WZflow_vector} is represented graphically. 
The circle with cross associated with the end points of the flow time zero is a one-point vertex defined by 
\begin{align}
  \begin{minipage}[c]{5cm}
    \centering
    \includegraphics[width=18mm]{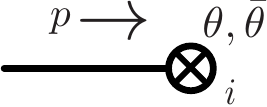}
  \end{minipage}
 \begin{minipage}[l]{5cm}
   \vspace{2mm}
    \centering
    $ \hspace{-30mm}= \hspace{8mm}{\bf \Phi}_{0,i}(p,\theta,\bar\theta)$.
  \end{minipage}\label{one-point-vertex}
  \end{align}
The flow vertex shown by an open circle is defined as
\begin{align}
  \begin{minipage}[c]{5cm}
    \centering
    \includegraphics[width=25mm]{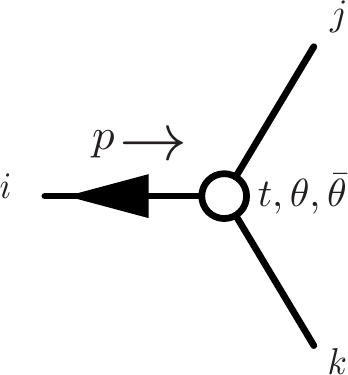}
  \end{minipage}
 \begin{minipage}[l]{5cm}
   \vspace{2mm}
    \centering
    $ \hspace{-20mm}= \hspace{8mm}g_{ijk}\bar{\bf \Delta}_{ii}(p,\theta,\bar\theta)$,
  \end{minipage}\label{flow_vertex}
  \end{align}
where an operator $ \bar{\bf\Delta}_{ii}(p,\theta,\bar\theta)$ acts upon the outgoing line with the index $i$. 
For each vertex (one-point and flow vertex), the Grassmann integral $\int d^2\theta d^2\bar\theta$ is performed. In addition, for the flow vertex, the flow time $t$ is integrated out from $0$ to $\infty$.

The flow line connecting the vertices  is defined by
\begin{align}
  \begin{minipage}[c]{4cm}
    \centering
    \includegraphics[width=35mm]{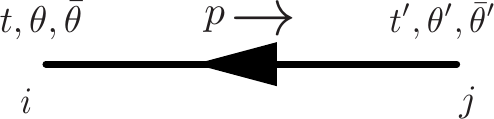}
  \end{minipage}
 \begin{minipage}[l]{8cm}
   \vspace{0mm}
    \centering
    $ \hspace{10mm}= \hspace{8mm}\Theta(t-t^\prime) {\bf \tilde K}_{t-t^\prime, ij}(p, \theta-\theta^\prime, \bar \theta-\bar\theta^\prime)$.
  \end{minipage}\label{flow_line}
  \end{align}
where ${\bf \tilde K}_{t}(p, \theta, \bar \theta)=\int d^4x e^{-ipx}{\bf K}_{t}(x, \theta, \bar \theta)$   
and $\Theta(t)$ is the Heaviside step function. The arrow indicates the direction of increasing flow time.

As for the momenta, at each flow vertex, the momentum conservation is assumed, 
and an undermined momentum of ingoing flow lines is integrated.

In Fig.~\ref{fig1}, the treelike graph begins at a single square of flow time $t$
and terminates at the one-point vertices of flow time $0$. The flow time runs from $0$ to $t$ keeping the time order with step functions. The initial condition (\ref{inital_condition_Phi}) tells  us that this iterative approximation 
may be understood as the perturbative expansion of one-third power of the coupling constant $g^{\frac{1}{3}}$.

%%%%%%%%%%%%%%%%%%%%%%%%%%%%%%%%%%%%%%%%%%
%
%
%            Section 4. 4
%
%
%%%%%%%%%%%%%%%%%%%%%
\subsection{Super-Feynman rules}\label{Feynman_rule}

We move on to perturbative calculations of correlation functions of ${\bf \Psi}_i$ combining the above iterative approximation of the Wess-Zumino flow and the super-Feynman rules in the Wess-Zumino model at $t=0$ discussed in Sec.~\ref{PT}. 

For example, the leading order contribution to the two-point function is diagrammatically represented as
\begin{align}
  \hspace{-15mm}\begin{minipage}[c]{10cm}
      \centering
\includegraphics [width=80mm]{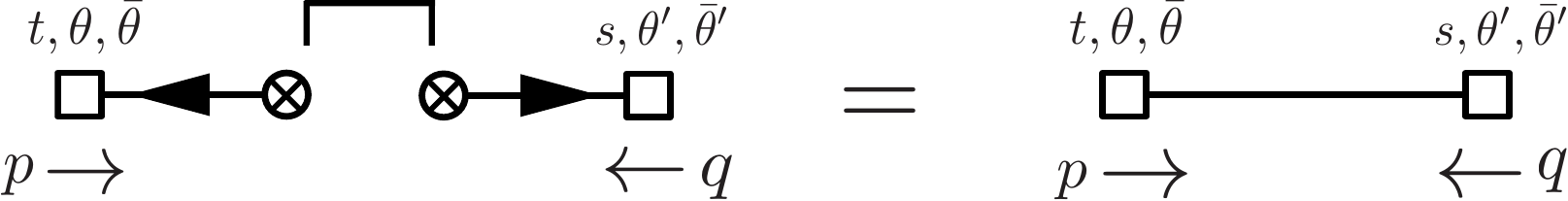}
  \end{minipage}\hspace{-5mm}.
\label{contraction_2Phi}
\end{align}
The staple symbol on the lhs denotes the contraction between two boundary fields ${\bf \Phi}_0$, which is given at the leading order as
\begin{align}
\begin{split}
&\la{\bf \Phi}_{0,i}(p,\theta,\bar\theta){\bf\Phi}_{0,j}(p^\prime,\theta^\prime,\bar\theta^\prime) \ra \\
&\qquad = 
g^{\frac{2}{3}}
{\bf D}_{ij}(p,\theta,\bar\theta) 
\ (2\pi)^4 \delta^4(p+p^\prime) 
\delta^2(\theta-\theta^\prime)\delta^2(\bar \theta -\bar \theta^\prime),
\end{split}
\label{Phi0_correlation_function}
\end{align}
where 
\begin{align}
{\bf D}(p,\theta,\bar\theta) = 
\frac{1}{\sqrt{p^2+m^2}}
\left(
\begin{array}{cc}
{\rm sin}\left(\beta_0(p) \right) \frac{\bar D^2}{4}  & 
{\rm cos}\left(\beta_0(p) \right) \frac{\bar D^2 D^2}{16 \sqrt{p^2}}  \\
{\rm cos}\left(\beta_0(p) \right) \frac{D^2 \bar D^2}{16 \sqrt{p^2}}  &
{\rm sin}\left(\beta_0(p) \right) \frac{D^2}{4}
\end{array}
\right)
\label{PhiPhi_correlation_function}
\end{align}
for ${\rm tan}(\beta_0(p))=m/\sqrt{p^2}$.

As shown in Eq. \eqref{contraction_2Phi}, 
we obtain the two-point function of $\bf \Psi$ at the leading order
taking a contraction between two $\bf \Phi_0$ for two tree-level solutions of $\bf \Psi$ as 
\begin{align}
\begin{split}
&\la{\bf \Psi}_i(t,p,\theta,\bar\theta){\bf\Psi}_j(s,q,\theta^\prime,\bar\theta^\prime) \ra\\
&\qquad = g^{\frac{2}{3}}
{\bf D}_{t+s, ij}(p,\theta,\bar\theta) 
\ (2\pi)^4 \delta^4(p+q) 
\delta^2(\theta-\theta^\prime)\delta^2(\bar \theta -\bar \theta^\prime),
\label{Psi_correlation_function}
\end{split}
\end{align}
where 
\begin{align}
{\bf D}_{t}(p,\theta,\bar\theta) = 
\frac{e^{-tp^2}}{\sqrt{p^2+m^2}}
\left(
\begin{array}{cc}
{\rm sin}\left(\beta_t(p) \right) \frac{\bar D^2}{4}  & 
{\rm cos}\left(\beta_t(p) \right) \frac{\bar D^2 D^2}{16 \sqrt{p^2}}  \\
{\rm cos}\left(\beta_t(p) \right) \frac{D^2 \bar D^2}{16 \sqrt{p^2}}  &
{\rm sin}\left(\beta_t(p) \right) \frac{D^2}{4}
\end{array}
\right)
\label{D_t}
\end{align}
for $\beta_t(p) = \beta_0(p) + t M \sqrt{p^2}$.

Thus, a field propagator associated with Eq.\eqref{Psi_correlation_function} is defined by
\begin{align}
  \begin{minipage}[c]{3cm}
    \centering
    \includegraphics[width=35mm]{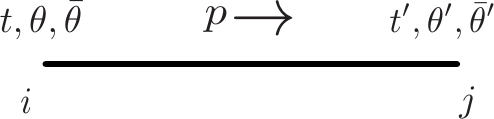}
  \end{minipage}
 \begin{minipage}[l]{9cm}
   \vspace{0mm}
    \centering
    $ \hspace{5mm}= \hspace{5mm}g^{\frac{2}{3}} {\bf D}_{t+t^\prime,ij}(p, \theta, \bar \theta)
    \delta^2(\theta-\theta^\prime)\delta^2(\bar \theta -\bar \theta^\prime)$.
  \end{minipage}\label{field_propagator}
  \end{align}
The time dependence appears as a sum of two boundary times, 
and the diagram of field propagator is shown by a line without an arrow.  
Since Eq.~\eqref{D_t} reproduces Eq.~\eqref{PhiPhi_correlation_function} for $t=0$, 
Eq.~\eqref{field_propagator} contains all of the field propagators 
such as $\la{\bf \Phi}_0{\bf\Phi}_0\ra$ and the mixed one $\la{\bf \Phi}_0{\bf\Psi}\ra$, as well as 
$\la{\bf \Psi\Psi\ra}$. 
Note that each field propagator is counted as $g^{\frac{2}{3}}$ in the perturbation theory.

We reformulate the perturbation theory at $t=0$ in terms of ${\bf \Phi}_0$ because the ${\bf\Phi}_0$ propagator is treated
uniformly with flow propagators.  Unlike the perturbation theory given in Sec.~\ref{PT}, the GRS propagator $\Delta_{GRS}$ is not used. 
The super-Feynman rules at $t=0$ should be modified to make fit with the rules for the iterative approximation of the Wess-Zumino flow equation. 
First, we rewrite the interaction part of the action (\ref{int_wz}) as 
\begin{align}
S_{int}=- \int d^8z \left\{ \frac{1}{3!} h_{ijk} \left(\frac{{\bf\Delta}}{\Box}{\bf\Phi}_{0}\right)_i{\bf\Phi}_{0,j}{\bf\Phi}_{0,k}\right\}(z),
\label{int_wz_mod}
\end{align}
where $h_{ijk}=2\delta_{ij}\delta_{ik}$.
The three-point vertex of the flow time zero may be defined by
\begin{align}
  \begin{minipage}[l]{5cm}
    \centering
    \includegraphics[width=25mm]{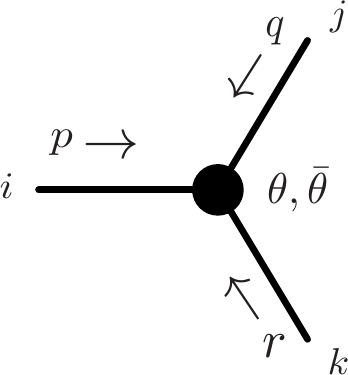}
  \end{minipage}
     \hspace{-0mm}= \hspace{8mm}-h_{ijk} \frac{{\bf\Delta}_{ii}(p,\theta,\bar\theta)}{p^2}, 
    \label{boundary_vertex}
\end{align}
where ${\bf\Delta}_{ii}(p,\theta,\bar\theta)$ acts on an internal line $p,i$. 
This is because
 ${\bf\Delta}_{ii}(p,\theta,\bar\theta)/p^2$ can be changed to 
${\bf\Delta}_{jj}(q,\theta,\bar\theta)/q^2$ or ${\bf\Delta}_{kk}(r,\theta,\bar\theta)/r^2$ 
by using the identity $\frac{\bar{\bf \Delta}{\bf \Delta }}{\Box}  {\bf \Phi}_0  =  {\bf \Phi}_0$ for Eq.\eqref{int_wz_mod}.
For each boundary vertex, the Grassmann integral $\int d^2\theta d^2\bar\theta$ is performed. 

Now, we consider the following one-loop correction to the two-point function, 
including one flow vertex (open circle) and one ordinary vertex (filled circle)\footnote{The boundary vertex attached to three ${\bf\Phi}_0$ is given by 
a product of Eqs.~\eqref{boundary_vertex} and \eqref{one-point-vertex}.}:
\begin{align}
  \hspace{-5mm}\begin{minipage}[c]{10cm}
      \centering
\includegraphics [width=120mm]{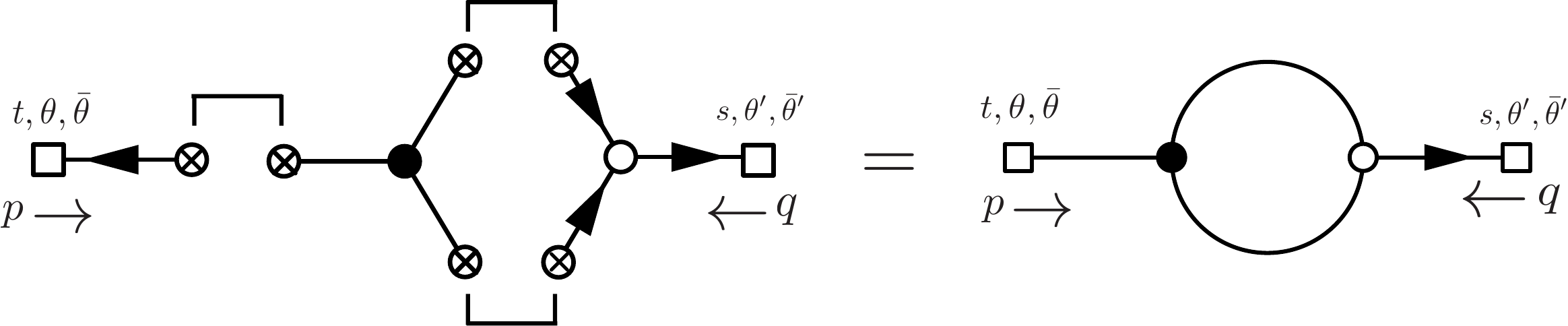}
  \end{minipage}\hspace{20mm}.
\label{contraction_2Phi_one_loop}
\end{align}
As in the tree-level case, performing the contraction between two ${\bf \Phi}_0$ yields a field propagator.  
In this case, the three lines without arrows on the rhs indicate the mixed propagators associated with $\la{\bf\Phi}_0{\bf\Psi}\ra$.

\begin{figure}[]\begin{center}
\includegraphics [width=120mm]
{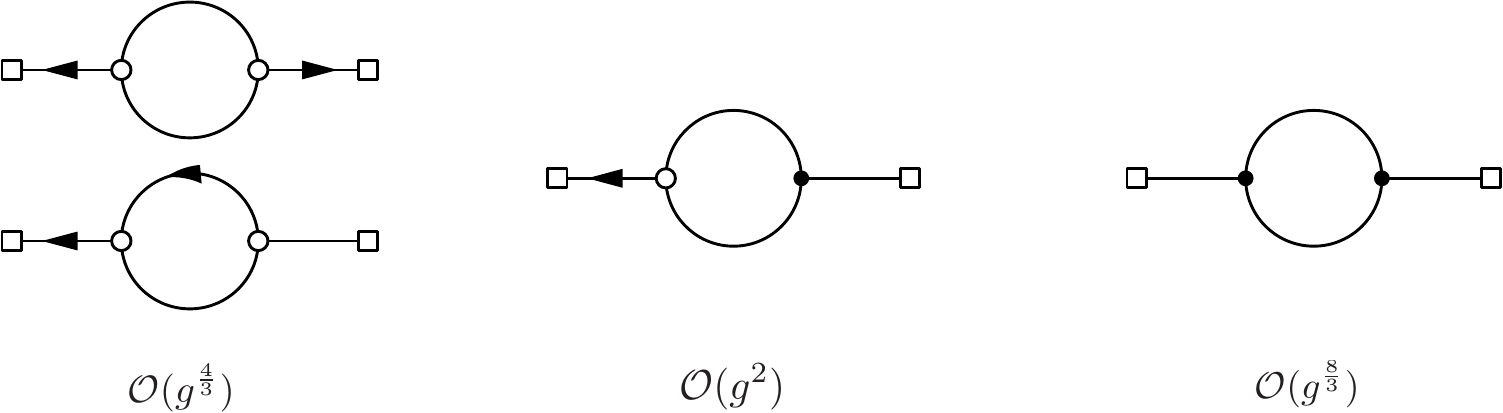}\end{center}
\caption{One-loop diagrams}
\label{oneloop}\end{figure}

Here, we mention that the coupling expansion does not naively  correspond to the loop expansion. 
This is because the $g$ dependence arises only from the field propagators of the order $g^{2/3}$, 
and the vertices and flow propagator do not depend on $g$.
Each one-loop diagram in Fig.~\ref{oneloop} has different orders $g^{2n/3}$ where $n$ is
the number of field propagators.

The super-Feynman rules for the correlation functions of ${\bf\Psi}$ in the momentum space are summarized as follows:
\begin{itemize}
\item[(a)] Use Eq.\eqref{flow_line} for a flow line that is an outgoing line emanated from each flow vertex.

\item[(b)] Use Eq.\eqref{field_propagator} for a field propagator by which two points (flow vertices, boundary vertices, 
and starting points denoted by $\square$) are connected.

\item[(c)] Use Eq.\eqref{flow_vertex} for each flow vertex, and
use Eq.\eqref{boundary_vertex} for each boundary vertex. 
For each flow vertex at $t$, perform the flow time integral $\int_{0}^{\infty}dt$.   
For all the flow and boundary vertices at $(\theta,\bar\theta)$, perform the Grassmann integral $\int d^2\theta  d^2\bar\theta$. 

\item[(d)] Impose the momentum conservation at each vertex and integrate over undetermined loop momenta.

\item[(e)] Compute the usual combinatoric factors.

\end{itemize}
These rules are given in Euclidean space. In addition, we mention rules and properties that are common with the Yang-Mills flow\cite{Luscher:2011bx}. 
Diagrams with closed flow line loops are absent because any loop has at least a field propagator. 
The flow lines depend on the difference between two flow times of end points.
The flow time dependence of propagators are determined by the sum of flow times at the end points. 
%%%%%%%%%%%%%%%%%%%%%

%%%%%%%%%%%%%

\subsection{The massive free flow}
\label{massivefreeflow}

We consider the massive free flow, dropping the interaction terms from the flow equations 
(but the boundary Wess-Zumino model has the interactions). 
The exact solution is
\begin{align}
{\bf \Psi}(t,z) = \int d^8 z^\prime {\bf K}_t (z-z^\prime) {\bf \Phi}_0(z^\prime).
\label{free_flow_solution}
\end{align}
Then, recalling the definition of ${\bf \Phi}_0$ (\ref{inital_condition_Phi}), a correlation function of the flowed fields $\langle {\bf \Psi}(t_1,z_1) {\bf \Psi}(t_2,z_2) \ldots {\bf \Psi}(t_n,z_n) \rangle$ can be given by a linear combination of correlation functions of the renormalized fields $\Phi_R(z_i)$ and $\bar \Phi_R(z_i)$
with $(g_R)^{\frac{n}{3}}$. 
In the renormalized perturbation theory, when evaluating the correlators of $\Phi_R(z_i)$ and $\bar \Phi_R(z_i)$, 
UV divergences are renormalized by the normal counterterms. 
So, in the case of the massive free flow, 
any correlation function of the flowed fields is UV finite for any nonzero flow time 
if the Wess-Zumino model is properly 
renormalized.

\subsection{Power counting theorem}\label{sec:power}
We can calculate the superficial degrees of divergence in the perturbation theory of the Wess-Zumino flow
using the super-Feynman rule given in Sec.~\ref{Feynman_rule}.

Since the field propagators given in Eq.~\eqref{field_propagator} have $t$-dependent functions,  
we need to evaluate 
the following integrals for each flow vertex: 
\begin{align}
I(p^2) \equiv \int_0^\infty dt e^{-tp^2}f(t,p^2),
\end{align}
where $p$ is a loop momentum and external momenta are set to zero for simplicity. 
After a short calculation, we find that, for large $p^2$,
\begin{align}
I(p^2) = p^{-2} f(0,p^2) + \left(p^{-2}\right)^2 f^{(1)} (0,p^2) + \cdots,
\label{t_int_identity}
\end{align}
where $f^{(n)} (t,p^2)= d^n f(t,p^2)/dt^n$. 
Since flow propagators with the same chirality and field propagators have 
$f(t,p^2)\sim {\rm cos}(tM\sqrt{p^2})$, ${\rm cos}(\beta_t(p^2))$, ${\rm sin}(\beta_t(p^2))$, 
the extra suppression factor appears as $p^{-2}$ from the first term of \eqref{t_int_identity}. 
Whereas, for massive flow propagators with the opposite chirality, 
$f(t,p^2)\sim {\rm sin}(tM\sqrt{p^2})$ leads to $f(0,p^2)=0$ 
and the extra factor becomes $p^{-3}$ from the second term of \eqref{t_int_identity}. 

At each flow vertex with an external flow line, we can apply the identity $\frac{\bar {\bf \Delta} {\bf \Delta}}{\Box} {\bf \Psi}={\bf \Psi}$ 
to an internal line and move a factor of $\bar {\bf \Delta}$ to the external line by integrals of parts. 
This transformation leads to an extra suppression factor $p^{-1}$ because a factor 
$\frac{{\bf \Delta}}{\Box} $ remains at the internal line. 
This type of transformation cannot be applied to the boundary vertex because,
since it is made of fields with the same chirality, the partial integration of $\bar {\bf \Delta}$ does not work.  

Consider a 1PI supergraph with $L$ loops, $V$ boundary vertices, $V_f$ flow vertices, 
$E$ external field lines, $E_f$ external flow lines, and $P$ field propagators,
of which $C$ are massive field propagators with the same chirality, $\Psi\Psi$ and $\bar\Psi\bar\Psi$, 
and $P_f$ flow propagators, of which $C_f$ are massive flow propagators with the opposite chirality. 
Each loop has a $d^4 p$ integral, 
and the identity $\delta_{12} D^2 \bar D^2 \delta_{12}= 16\delta_{12}$ still applies in this case
to remove a $D^2\bar D^2 \sim p^2$ at each loop.  
At $t=0$, 
$\Psi\bar\Psi$ propagators behave as $p^0$, while massive chiral propagators behave as $p^{-1}$ for large $p^2$. 
We have extra suppression factors $p^{-2}$ from the boundary of  $t$ integrations at each flow vertex
discussed above. 
For massive flow propagators with the opposite chirality, we have $p^{-3}$ instead of $p^{-2}$. 
Each boundary vertex has a factor of ${\bf \Delta}_{ii}/p^2 \sim p^{-1}$ on one of the internal lines.
Each {\it internal} outgoing flow line emanated from the flow vertex has a factor of $\bar{\bf \Delta}_{ii} \sim p$.
Each external flow line has a suppression factor $p^{-1}$ from the discussion using 
the identity $\frac{\bar {\bf \Delta} {\bf \Delta}}{\Box} {\bf \Psi}={\bf \Psi}$. 

Thus, we find that the superficial degrees of divergence $d$ is given by
\begin{align}
d=2L-C-2V_f-C_f-V+V_f-E_f -E_f \label{sdod}. 
\end{align}
Using a topology relation $L-P-P_f+V+V_f=1$ and a few relations
such as $3V+3V_f = E+E_f+2P+2P_f$ (each vertex has three lines) and 
$V_f=E_f+P_f$ (the flow vertex has an outgoing flow line), where 
$E_f \ge 1$ for nonzero $V_f$,
we finally obtain 
\begin{align}
d=2-C-C_f-E-3E_f.
\end{align}
This  shows that any super-Feynman graph with flow vertices is UV finite at all orders of perturbation theory. 
The remaining divergences for $V_f=E_f=0$ arise from boundary vertices and cancel as
in the massive free flow case because $n$-point functions of ${\bf \Psi}(t,z)$ 
are those of ${\bf K}_t {\bf \Phi}_0(z)$ for $V_f=0$ and 
${\bf \Phi}_0$ is given by $g_R$ and renormalized fields $\Phi_R$ from \eqref{inital_condition_Phi}.
 We can conclude that any correlation function of flowed fields 
is UV finite in the Wess-Zumino flow at all orders of perturbation theory.

\section{Summary} 
\label{sec:summary}

We introduced a supersymmetric gradient flow with renormalization-invariant couplings in the Wess-Zumino model and showed that correlation functions of the flowed superfield are UV finite using a power counting theorem for 1PI supergraphs based on super-Feynman rules.  In particular, we found that the interaction terms of the flow equation do not contribute to divergent graphs, only terms of the boundary theory do. After the parameter renormalization in the boundary theory, the remaining divergence of the wave function can be removed by taking initial conditions to be renormalization invariant. Thus, we found that any correlation function of the flowed superfield is UV finite at all orders of the perturbation theory.

In nonsupersymmetric scalar field theory, including the mass term and a term like $\phi^4$ interaction yields nonremovable divergences. Even in the massless free flow, a wave function renormalization remains. Some kind of symmetry could be necessary for the UV finiteness property. In the Yang-Mills flow, the BRS symmetry guarantees the UV finiteness, whereas in the Wess-Zumino flow the supersymmetry plays a crucial role to hold the property in a mechanism that is quite different from the Yang-Mills flow. 
  
The existence of the nonrenormalization theorem is significant in our proof because it leads to the renormalization-invariant initial condition [Eqs.~\eref{inital_condition_Psi} and \eref{inital_condition_Phi}] and the invariant mass [Eq.~\eref{inv_mass}] in the Wess-Zumino flow [Eq.~\eref{WZflow_vector}]. The UV finiteness is a direct consequence of these invariances. Therefore, it is unclear whether our results can be extended to other theories that do not have a nonrenormalization theorem.

Without the renormalization-invariant initial conditions \eref{inital_condition_Psi} and \eref{inital_condition_Phi}, the wave function renormalization remains, and the extra wave function renormalization of the flowed superfield, as in gradient flow of quark fields, makes the correlation function finite. On the other hand, if the flow equations are not given by renormalization-invariant coupling constants, the perturbative renormalizability breaks down completely.

Gradient flows have been  successfully applied to various research such as nonperturbative renormalization group, holographic descriptions of field theory, and lattice simulations. In addition, supersymmetry has been actively  studied in particle physics in a variety of ways. Therefore, supersymmetric gradient flows can be  expected to have various applications. The techniques developed in this article will be very useful for subsequent studies using supersymmetric gradient flows.

\section*{Acknowledgement}

This work was supported by JSPS KAKENHI Grants No.~18K13546, No.~19K03853, No.~20K03924, No.~21K03537, and No.~22H01222.

\appendix

\section{Convention in Euclidean space}
\label{notation}

In order to obtain the Euclidean theory from the Minkowski one with metric $\eta_{\mu\nu}={\rm diag}\{-1,1,1,1\}$ in Ref.~\cite{Wess:1992cp}, we use the Wick rotation $x_0 \rightarrow -i x_0$  to move on to the Euclidean signature. 
The Euclidean four-dimensional $\sigma$ matrices are defined 
as $\sigma_0 =\bar\sigma_0 \equiv -i\mathbf{1}$; the others are the same. 
The auxiliary fields in the chiral superfields are replaced as $F, F^* \rightarrow iF, iF^*$.  
The Euclidean action is defined as $S_{E} =-i S $ after the Wick rotation.  

The Fourier transformation is defined by 
\begin{align}
\Phi(x,\theta,\bar\theta)=\int \frac{d^4p}{(2\pi)^4} e^{ipx}\tilde{\Phi}(p,\theta,\bar\theta).
\end{align}

\subsection{Spinors and $\sigma$ matrices}
\label{sec:notation_alg}

Let $\psi_\alpha$ ($\alpha=1,2$) be a $SU(2)_R$ spinor and $\bar\psi_{\dot\alpha}$  ($\dot\alpha=1,2$) be a $SU(2)_L$ spinor, then they are not related to each other under the complex conjugate in the four-dimensional Euclidian space. 
We define  the invariant tensors of  $SU(2)_R$ and $SU(2)_L$ as 
\begin{align}
 \epsilon_{21}=\epsilon^{12}=\epsilon_{\dot2\dot1}=\epsilon^{\dot1\dot2}=1,\quad
  \epsilon_{12}=\epsilon^{21}=\epsilon_{\dot1\dot2}=\epsilon^{\dot2\dot1}=-1
\end{align}
with the others being zero, so that 
$\epsilon_{\alpha\beta}\epsilon^{\beta\gamma}=\delta_{\alpha}{}^{\gamma}$ and  
$\epsilon_{\dot\alpha\dot\beta}\epsilon^{\dot\beta\dot\gamma}=\delta_{\dot\alpha}{}^{\dot\gamma}$.
Then spinors with upper and lower indices are related through the invariant tensors,
\begin{align}
\psi^\alpha = \epsilon^{\alpha\beta}\psi_\beta,\quad
\psi_\alpha = \epsilon_{\alpha\beta}\psi^\beta.
\end{align}
We use the following spinor summation convention:
\begin{align}
\psi \chi = \psi^\alpha \chi_\alpha, 
\qquad  
\bar\psi \bar\chi =  \bar\psi_{\dot\alpha} \bar\chi^{\dot\alpha}.
\label{contraction}
\end{align}

The $\sigma$ matrices in the Euclidean space $(\sigma_{\mu})_{\alpha\dot{\beta}}$ and $(\bar\sigma_{\mu})^{\dot\alpha \beta}$ 
are defined as
\begin{align}
& \sigma_0 =
\left(
\begin{array}{cc} -i &  0    \\  0 &  -i   \end{array}
\right),
\qquad  
 \sigma_1 =
\left(
\begin{array}{cc}  0 &  1    \\  1 &  0   \end{array}
\right),  \nonumber \\ 
& \sigma_2 =
\left(
\begin{array}{cc}  0 &  -i    \\  i &  0   \end{array}
\right),
\qquad  
 \sigma_3 =
\left(
\begin{array}{cc}  1 &  0    \\  0 &  -1   \end{array}
\right),
 \\
& (\bar{\sigma}_{\mu})^{\dot{\alpha}\alpha} = \epsilon^{\dot\alpha\dot\beta}\epsilon^{\alpha\beta}
 (\sigma_{\mu})_{\beta\dot{\beta}},
\nonumber \\
& \bar \sigma_0 = \sigma_0, \qquad \bar\sigma_i=-\sigma_i \ \ (i=1,2,3). \nonumber  
\end{align}
For more detail on the spinor algebra, see Ref.~\cite{Kadoh:2018qwg}.

%
%
%   Chiral superfield
%
%
\subsection{Chiral superfield}

The supercharges $Q_\alpha$ and $\bar Q_{\dot\alpha}$ are defined as difference operators on the superspace labeled by $z=(x_\mu,\theta_\alpha,\bar\theta_{\dot\alpha})$,
\begin{align}
\begin{split}
& Q_{\alpha} = 
{\partial_{\alpha}} 
-i(\sigma_{\mu})_{\alpha\dot{\alpha}}\bar\theta^{\dot\alpha}\partial_{\mu} \\
& \bar{Q}_{\dot\alpha} = 
-{\partial_{\dot\alpha}} 
+i\theta^{\alpha}(\sigma_{\mu})_{\alpha\dot{\alpha}}\partial_{\mu},
\end{split}
\end{align}
where $\partial_\alpha = \frac{\partial}{\partial\theta^{\alpha}},\ 
 \partial_{\dot\alpha} = \frac{\partial}{\partial\bar\theta^{\dot\alpha}}$, and
 $\partial_{\mu} = \frac{\partial}{\partial x_\mu}$.
The associated supercovariant derivatives that commute with the supercharges are defined as
\begin{align}
\begin{split}
& D_{\alpha} =
 {\partial_{\alpha}}
+i(\sigma_{\mu})_{\alpha\dot{\alpha}}\bar\theta^{\dot\alpha}\partial_{\mu}  \\
& \bar{D}_{\dot\alpha} =
 -{\partial_{\dot\alpha}}
-i\theta^{\alpha}(\sigma_{\mu})_{\alpha\dot{\alpha}}\partial_{\mu}.
\end{split}
\end{align}
These difference operators obey 
\begin{align}
\begin{split}
& \{Q_{\alpha},\bar{Q}_{\dot{\alpha}}\} = 2i(\sigma_{\mu})_{\alpha\dot{\alpha}}\partial_{\mu}  \\
& \{D_{\alpha},\bar{D}_{\dot{\alpha}}\} = -2i(\sigma_{\mu})_{\alpha\dot{\alpha}}\partial_{\mu}
\end{split}
\end{align}
and the other anticommutation relations vanish.
After a short calculation, one can show the useful identities 
\begin{align}
\begin{split}
D^2\bar{D}^2D^2=&16\Box D^2,\\
\bar{D}^2D^2\bar{D}^2=&16\Box\bar{D}^2,
\label{identity_Box}
\end{split}
\end{align}
where $\Box=\partial_\mu\partial_\mu$.

The chiral and antichiral superfields $\Phi(x,\theta,\bar\theta)$ and $\bar\Phi(x,\theta,\bar\theta)$ are characterized by the constraints  $\bar D_{\dot\alpha}\Phi = 0$ and $D_{\alpha} \bar\Phi =0$, 
respectively. They are expanded in $\theta$ and $\bar\theta$ as
\begin{align}
\Phi(x,\theta,\bar\theta) =&
 A(x) +i\theta\sigma_{\mu}\bar\theta \partial_{\mu} A(x) +\frac{1}{4}\theta \theta \bar\theta \bar\theta \Box A(x) \nonumber\\
 & +\sqrt{2}\theta\psi(x) -\frac{i}{\sqrt{2}}\theta\theta \partial_{\mu}\psi(x)\sigma_{\mu}\bar\theta 
                          +i\theta\theta F(x), \\
\bar\Phi(x,\theta,\bar\theta) =&
 A^*(x) -i\theta\sigma_{\mu}\bar\theta \partial_{\mu} A^*(x) +\frac{1}{4}\theta \theta \bar\theta \bar\theta \Box A^*(x) \nonumber\\
 & +\sqrt{2}\bar\theta \bar\psi(x) +\frac{i}{\sqrt{2}}\bar\theta\bar\theta \theta \sigma_{\mu}\partial_{\mu}\bar\psi(x) 
                          +i\bar\theta\bar\theta F^*(x),
\end{align}
where $A$ and $F$ are complex bosonic fields, and $\psi, \bar\psi$ are two component spinors. 
Note that $\bar \Phi$ is not a complex conjugate of $\Phi$ in this 
Euclidean theory.
One can easily find the following projection operators for the chiral superfields:
\begin{align}
\frac{\bar{D}^2D^2}{16\Box}\Phi=\Phi,\label{identity_Box1}\\
\frac{D^2\bar{D}^2}{16\Box}\bar\Phi=\bar\Phi.
\label{identity_Box2}
\end{align}

Introducing new coordinate $(y, \theta, \bar\theta)$ with $y_{\mu}=x_{\mu}+i\theta\sigma_{\mu}\bar\theta$, 
the derivative operators and $\Phi$ are expressed as 
\begin{align}
\begin{split}
& Q_{\alpha} = {\partial_{\alpha}}, \\
& \bar{Q}_{\dot\alpha} = 
-{\partial_{\dot\alpha}} 
+2i\theta^{\alpha}(\sigma_{\mu})_{\alpha\dot{\alpha}}\partial_{\mu},  \\
& D_{\alpha} =
 {\partial_{\alpha}}
+2i(\sigma_{\mu})_{\alpha\dot{\alpha}}\bar\theta^{\dot\alpha}\partial_{\mu}, \\
& \bar{D}_{\dot\alpha} =
 -{\partial_{\dot\alpha}}, \\
& \Phi(y,\theta) = A(y)+\sqrt{2}\theta\psi(y)+i\theta\theta F(y),
\end{split}
\end{align}
while in $(\bar{y}, \theta, \bar\theta)$ with $\bar{y}_{\mu}=x_{\mu}-i\theta\sigma_{\mu}\bar\theta$,  
\begin{align}
\begin{split}
& Q_{\alpha} = 
{\partial_{\alpha}} 
-2i(\sigma_{\mu})_{\alpha\dot{\alpha}}\bar\theta^{\dot\alpha}\partial_{\mu},  \\
& \bar{Q}_{\dot\alpha} = 
-{\partial_{\dot\alpha}},  \\
& D_{\alpha} =
 {\partial_{\alpha}},  \\
& \bar{D}_{\dot\alpha} =
 -{\partial_{\dot\alpha}}  
 -2i\theta^{\alpha}(\sigma_{\mu})_{\alpha\dot{\alpha}}\partial_{\mu},  \\
& \bar\Phi(\bar{y},\bar\theta) = A^*(\bar{y})+\sqrt{2}\bar\theta\bar\psi(\bar{y})+i\bar\theta\bar\theta F^*(\bar{y}).
\end{split}
\end{align}
Note that $\bar y$ is not a complex conjugate of $y$ in the Euclidean space.

\subsection{Integral and delta function over Grassmann coordinates}
The volume element of the superspace $z=(x_\mu,\theta_{\alpha},\bar\theta_{\dot\alpha})$ is
\begin{align}
 d^8z = d^4x\, d^2\theta\, d^2\bar\theta,
\end{align}
where 
\begin{align}
 \int d^2\theta \, \theta^2 = 1,
 \quad
 \int d^2\bar\theta \, \bar\theta^2 = 1. 
\label{theta_integral}
\end{align}
Under the Euclidean space integral, the Grassmann integrals can be interpreted as
\begin{align}
 \int d^4x\, d^2\theta = \int d^4x\, \left(-\frac{D^2}{4}\right), 
 \quad
 \int d^4x\, d^2\bar\theta = \int d^4x\, \left(-\frac{\bar D^2}{4}\right)
\end{align}
and 
\begin{align}
 \int d^4x\, d^2\theta \, d^2\bar\theta = \int d^4x\, \left(\frac{D^2 \bar D^2}{16}\right).
\end{align}

The delta functions are defined as
\begin{align}
 \delta^2(\theta) = \theta^2,
 \quad
 \delta^2(\bar\theta) = \bar\theta^2,
\end{align}
such that 
\begin{align}
 \int d^2\theta\, \delta^2(\theta) = 1,
 \quad
 \int d^2\bar\theta\, \delta^2(\bar\theta) = 1. 
\end{align}

The functional derivatives of chiral superfields $\Phi(z)$  and  $\bar{\Phi}(z)$ are
\begin{align}
\frac{\delta \Phi(z_1)}{\delta \Phi(z_2)}
=-\frac{\bar{D}_1^2}{4}\delta^8 (z_1-z_2),\\
\frac{\delta \bar{\Phi}(z_1)}{\delta \bar{\Phi}(z_2)}
=-\frac{D_1^2}{4}\delta^8 (z_1-z_2),
\end{align}
where 
\begin{align}
\delta^8 (z_1-z_2) = 
\delta^4(x_1-x_2)\delta^2(\theta_1-\theta_2)\delta^2(\bar\theta_1-\bar\theta_2).
\end{align}
We use the abbreviation $\delta_{12}=\delta^2(\theta_1-\theta_2)\delta^2(\bar\theta_1-\bar\theta_2)$ for simplicity.  
The following relation 
\begin{align}
\delta_{12}D^2\bar{D}^2\delta_{21}=\delta_{12}\bar{D}^2D^2\delta_{21}=16\delta_{12}
\end{align}
is useful in perturbative calculations.

\bibliographystyle{utphys}

\begin{thebibliography}{10}

\bibitem{Luscher:2010iy}
M.~Lüscher, ``{Properties and uses of the Wilson flow in lattice QCD},''
  \href{http://dx.doi.org/10.1007/JHEP08(2010)071,
  10.1007/JHEP03(2014)092}{{\em JHEP} {\bfseries 08} (2010) 071},
  \href{http://arxiv.org/abs/1006.4518}{{\ttfamily arXiv:1006.4518 [hep-lat]}}.
[Erratum: JHEP03,092(2014)].
%%CITATION = ARXIV:1006.4518;%%.

\bibitem{Luscher:2011bx}
M.~Luscher and P.~Weisz, ``{Perturbative analysis of the gradient flow in
  non-abelian gauge theories},''
  \href{http://dx.doi.org/10.1007/JHEP02(2011)051}{{\em JHEP} {\bfseries 02}
  (2011) 051},
\href{http://arxiv.org/abs/1101.0963}{{\ttfamily arXiv:1101.0963 [hep-th]}}.
%%CITATION = ARXIV:1101.0963;%%.

\bibitem{Yamamura:2015kva}
R.~Yamamura, ``{The Yang-Mills gradient flow and lattice effective action},''
  \href{http://dx.doi.org/10.1093/ptep/ptw097}{{\em PTEP} {\bfseries 2016}
  no.~7, (2016) 073B10},
\href{http://arxiv.org/abs/1510.08208}{{\ttfamily arXiv:1510.08208 [hep-lat]}}.
%%CITATION = ARXIV:1510.08208;%%.

\bibitem{Makino:2018rys}
H.~Makino, O.~Morikawa, and H.~Suzuki, ``{Gradient flow and the Wilsonian
  renormalization group flow},''
  \href{http://dx.doi.org/10.1093/ptep/pty050}{{\em PTEP} {\bfseries 2018}
  no.~5, (2018) 053B02},
\href{http://arxiv.org/abs/1802.07897}{{\ttfamily arXiv:1802.07897 [hep-th]}}.
%%CITATION = ARXIV:1802.07897;%%.

\bibitem{Abe:2018zdc}
Y.~Abe and M.~Fukuma, ``{Gradient flow and the renormalization group},''
  \href{http://dx.doi.org/10.1093/ptep/pty081}{{\em PTEP} {\bfseries 2018}
  no.~8, (2018) 083B02},
\href{http://arxiv.org/abs/1805.12094}{{\ttfamily arXiv:1805.12094 [hep-th]}}.
%%CITATION = ARXIV:1805.12094;%%.

\bibitem{Carosso:2018bmz}
A.~Carosso, A.~Hasenfratz, and E.~T. Neil, ``{Nonperturbative Renormalization
  of Operators in Near-Conformal Systems Using Gradient Flows},''
  \href{http://dx.doi.org/10.1103/PhysRevLett.121.201601}{{\em Phys. Rev.
  Lett.} {\bfseries 121} no.~20, (2018) 201601},
\href{http://arxiv.org/abs/1806.01385}{{\ttfamily arXiv:1806.01385 [hep-lat]}}.
%%CITATION = ARXIV:1806.01385;%%.

\bibitem{Sonoda:2019ibh}
H.~Sonoda and H.~Suzuki, ``{Derivation of a gradient flow from the exact
  renormalization group},'' \href{http://dx.doi.org/10.1093/ptep/ptz020}{{\em
  PTEP} {\bfseries 2019} no.~3, (2019) 033B05},
\href{http://arxiv.org/abs/1901.05169}{{\ttfamily arXiv:1901.05169 [hep-th]}}.
%%CITATION = ARXIV:1901.05169;%%.

\bibitem{Sonoda:2020vut}
H.~Sonoda and H.~Suzuki, ``{Gradient flow exact renormalization group},''
  \href{http://dx.doi.org/10.1093/ptep/ptab006}{{\em PTEP} {\bfseries 2021}
  no.~2, (2021) 023B05}, \href{http://arxiv.org/abs/2012.03568}{{\ttfamily
  arXiv:2012.03568 [hep-th]}}.

\bibitem{Miyakawa:2021hcx}
Y.~Miyakawa and H.~Suzuki, ``{Gradient flow exact renormalization group:
  Inclusion of fermion fields},''
  \href{http://dx.doi.org/10.1093/ptep/ptab100}{{\em PTEP} {\bfseries 2021}
  no.~8, (2021) 083B04}, \href{http://arxiv.org/abs/2106.11142}{{\ttfamily
  arXiv:2106.11142 [hep-th]}}.

\bibitem{Miyakawa:2021wus}
Y.~Miyakawa, H.~Sonoda, and H.~Suzuki, ``{Manifestly gauge invariant exact
  renormalization group for quantum electrodynamics},''
  \href{http://dx.doi.org/10.1093/ptep/ptac003}{{\em PTEP} {\bfseries 2022}
  no.~2, (2022) 023B02}, \href{http://arxiv.org/abs/2111.15529}{{\ttfamily
  arXiv:2111.15529 [hep-th]}}.

\bibitem{Miyakawa:2022qbz}
Y.~Miyakawa, ``{Axial anomaly in the gradient flow exact renormalization
  group},'' \href{http://arxiv.org/abs/2201.08181}{{\ttfamily arXiv:2201.08181
  [hep-th]}}.

\bibitem{Sonoda:2022fmk}
H.~Sonoda and H.~Suzuki, ``{One-particle irreducible Wilson action in the
  gradient flow exact renormalization group formalism},''
  \href{http://dx.doi.org/10.1093/ptep/ptac047}{{\em PTEP} {\bfseries 2022}
  no.~5, (2022) 053B01}, \href{http://arxiv.org/abs/2201.04448}{{\ttfamily
  arXiv:2201.04448 [hep-th]}}.

\bibitem{Hasenfratz:2022wll}
A.~Hasenfratz, C.~J. Monahan, M.~D. Rizik, A.~Shindler, and O.~Witzel, ``{A
  novel nonperturbative renormalization scheme for local operators},''
  \href{http://dx.doi.org/10.22323/1.396.0155}{{\em PoS} {\bfseries
  LATTICE2021} (2022) 155}, \href{http://arxiv.org/abs/2201.09740}{{\ttfamily
  arXiv:2201.09740 [hep-lat]}}.

\bibitem{Abe:2022smm}
Y.~Abe, Y.~Hamada, and J.~Haruna, ``{Fixed point structure of the gradient flow
  exact renormalization group for scalar field theories},''
  \href{http://dx.doi.org/10.1093/ptep/ptac021}{{\em PTEP} {\bfseries 2022}
  no.~3, (2022) 033B03}, \href{http://arxiv.org/abs/2201.04111}{{\ttfamily
  arXiv:2201.04111 [hep-th]}}.

\bibitem{Aoki:2015dla}
S.~Aoki, K.~Kikuchi, and T.~Onogi, ``{Geometries from field theories},''
  \href{http://dx.doi.org/10.1093/ptep/ptv131}{{\em PTEP} {\bfseries 2015}
  no.~10, (2015) 101B01},
\href{http://arxiv.org/abs/1505.00131}{{\ttfamily arXiv:1505.00131 [hep-th]}}.
%%CITATION = ARXIV:1505.00131;%%.

\bibitem{Aoki:2016ohw}
S.~Aoki, J.~Balog, T.~Onogi, and P.~Weisz, ``{Flow equation for the large $N$
  scalar model and induced geometries},''
  \href{http://dx.doi.org/10.1093/ptep/ptw106}{{\em PTEP} {\bfseries 2016}
  no.~8, (2016) 083B04},
\href{http://arxiv.org/abs/1605.02413}{{\ttfamily arXiv:1605.02413 [hep-th]}}.
%%CITATION = ARXIV:1605.02413;%%.

\bibitem{Aoki:2017uce}
S.~Aoki and S.~Yokoyama, ``{AdS geometry from CFT on a general conformally flat
  manifold},'' \href{http://dx.doi.org/10.1016/j.nuclphysb.2018.06.004}{{\em
  Nucl. Phys.} {\bfseries B933} (2018) 262--274},
\href{http://arxiv.org/abs/1709.07281}{{\ttfamily arXiv:1709.07281 [hep-th]}}.
%%CITATION = ARXIV:1709.07281;%%.

\bibitem{Aoki:2017bru}
S.~Aoki and S.~Yokoyama, ``{Flow equation, conformal symmetry, and anti-de
  Sitter geometry},'' \href{http://dx.doi.org/10.1093/ptep/pty013}{{\em PTEP}
  {\bfseries 2018} no.~3, (2018) 031B01},
\href{http://arxiv.org/abs/1707.03982}{{\ttfamily arXiv:1707.03982 [hep-th]}}.
%%CITATION = ARXIV:1707.03982;%%.

\bibitem{Aoki:2018dmc}
S.~Aoki, J.~Balog, and S.~Yokoyama, ``{Holographic computation of quantum
  corrections to the bulk cosmological constant},''
  \href{http://dx.doi.org/10.1093/ptep/ptz026}{{\em PTEP} {\bfseries 2019}
  no.~4, (2019) 043B06}, \href{http://arxiv.org/abs/1804.04636}{{\ttfamily
  arXiv:1804.04636 [hep-th]}}.

\bibitem{Aoki:2019bfb}
S.~Aoki, S.~Yokoyama, and K.~Yoshida, ``{Holographic geometry for
  nonrelativistic systems emerging from generalized flow equations},''
  \href{http://dx.doi.org/10.1103/PhysRevD.99.126002}{{\em Phys. Rev. D}
  {\bfseries 99} no.~12, (2019) 126002},
  \href{http://arxiv.org/abs/1902.02578}{{\ttfamily arXiv:1902.02578
  [hep-th]}}.

\bibitem{Aoki:2022lye}
S.~Aoki, J.~Balog, T.~Onogi, and S.~Yokoyama, ``{Special flow equation~and the
  GKP\textendash{}Witten relation},''
  \href{http://dx.doi.org/10.1093/ptep/ptad002}{{\em PTEP} {\bfseries 2023}
  no.~1, (2023) 013B03}, \href{http://arxiv.org/abs/2204.06855}{{\ttfamily
  arXiv:2204.06855 [hep-th]}}.

\bibitem{Makino:2014cxa}
H.~Makino, F.~Sugino, and H.~Suzuki, ``{Large-$N$ limit of the gradient flow in
  the 2D $O(N)$ nonlinear sigma model},''
  \href{http://dx.doi.org/10.1093/ptep/ptv044}{{\em PTEP} {\bfseries 2015}
  no.~4, (2015) 043B07},
\href{http://arxiv.org/abs/1412.8218}{{\ttfamily arXiv:1412.8218 [hep-lat]}}.
%%CITATION = ARXIV:1412.8218;%%.

\bibitem{Aoki:2014dxa}
S.~Aoki, K.~Kikuchi, and T.~Onogi, ``{Gradient Flow of O(N) nonlinear sigma
  model at large N},'' \href{http://dx.doi.org/10.1007/JHEP04(2015)156}{{\em
  JHEP} {\bfseries 04} (2015) 156},
\href{http://arxiv.org/abs/1412.8249}{{\ttfamily arXiv:1412.8249 [hep-th]}}.
%%CITATION = ARXIV:1412.8249;%%.

\bibitem{Makino:2014sta}
H.~Makino and H.~Suzuki, ``{Renormalizability of the gradient flow in the 2D
  $O(N)$ non-linear sigma model},''
  \href{http://dx.doi.org/10.1093/ptep/ptv028}{{\em PTEP} {\bfseries 2015}
  no.~3, (2015) 033B08},
\href{http://arxiv.org/abs/1410.7538}{{\ttfamily arXiv:1410.7538 [hep-lat]}}.
%%CITATION = ARXIV:1410.7538;%%.

\bibitem{Aoki:2016env}
S.~Aoki, J.~Balog, T.~Onogi, and P.~Weisz, ``{Flow equation for the scalar
  model in the large $N$ expansion and its applications},''
  \href{http://dx.doi.org/10.1093/ptep/ptx025}{{\em PTEP} {\bfseries 2017}
  no.~4, (2017) 043B01},
\href{http://arxiv.org/abs/1701.00046}{{\ttfamily arXiv:1701.00046 [hep-th]}}.
%%CITATION = ARXIV:1701.00046;%%.

\bibitem{Nakazawa:2003zf}
N.~Nakazawa, ``{N=1 superYang-Mills theory in Ito calculus},''
  \href{http://dx.doi.org/10.1143/PTP.110.1117}{{\em Prog. Theor. Phys.}
  {\bfseries 110} (2004) 1117--1150},
\href{http://arxiv.org/abs/hep-th/0302138}{{\ttfamily arXiv:hep-th/0302138
  [hep-th]}}.
%%CITATION = HEP-TH/0302138;%%.

\bibitem{Nakazawa:2003tz}
N.~Nakazawa, ``{Stochastic gauge fixing in N=1 supersymmetric Yang-Mills
  theory},'' \href{http://dx.doi.org/10.1143/PTP.116.883}{{\em Prog. Theor.
  Phys.} {\bfseries 116} (2007) 883--917},
\href{http://arxiv.org/abs/hep-th/0308081}{{\ttfamily arXiv:hep-th/0308081
  [hep-th]}}.
%%CITATION = HEP-TH/0308081;%%.

\bibitem{Kikuchi:2014rla}
K.~Kikuchi and T.~Onogi, ``{Generalized Gradient Flow Equation and Its
  Application to Super Yang-Mills Theory},''
  \href{http://dx.doi.org/10.1007/JHEP11(2014)094}{{\em JHEP} {\bfseries 11}
  (2014) 094},
\href{http://arxiv.org/abs/1408.2185}{{\ttfamily arXiv:1408.2185 [hep-th]}}.
%%CITATION = ARXIV:1408.2185;%%.

\bibitem{Aoki:2017iwi}
S.~Aoki, K.~Kikuchi, and T.~Onogi, ``{Flow equation of $\mathcal{N}$ = 1
  supersymmetric O(N) nonlinear sigma model in two dimensions},''
  \href{http://dx.doi.org/10.1007/JHEP02(2018)128}{{\em JHEP} {\bfseries 02}
  (2018) 128},
\href{http://arxiv.org/abs/1704.03717}{{\ttfamily arXiv:1704.03717 [hep-th]}}.
%%CITATION = ARXIV:1704.03717;%%.

\bibitem{Kadoh:2018qwg}
D.~Kadoh and N.~Ukita, ``{Supersymmetric gradient flow in $\mathcal{N}=1$
  SYM},'' \href{http://dx.doi.org/10.1140/epjc/s10052-022-10404-y}{{\em Eur.
  Phys. J. C} {\bfseries 82} no.~5, (2022) 435},
  \href{http://arxiv.org/abs/1812.02351}{{\ttfamily arXiv:1812.02351
  [hep-th]}}.

\bibitem{Kadoh:2019glu}
D.~Kadoh, K.~Kikuchi, and N.~Ukita, ``{Supersymmetric gradient flow in the
  Wess-Zumino model},''
  \href{http://dx.doi.org/10.1103/PhysRevD.100.014501}{{\em Phys. Rev.}
  {\bfseries D100} no.~1, (2019) 014501},
\href{http://arxiv.org/abs/1904.06582}{{\ttfamily arXiv:1904.06582 [hep-th]}}.
%%CITATION = ARXIV:1904.06582;%%.

\bibitem{Kadoh:2019flv}
D.~Kadoh and N.~Ukita, ``{Gradient flow equation in SQCD},''
  \href{http://dx.doi.org/10.22323/1.363.0199}{{\em PoS} {\bfseries
  LATTICE2019} (2020) 199}, \href{http://arxiv.org/abs/1912.13247}{{\ttfamily
  arXiv:1912.13247 [hep-lat]}}.

\bibitem{Bergner:2019dim}
G.~Bergner, C.~L\'opez, and S.~Piemonte, ``{Study of center and chiral symmetry
  realization in thermal $\mathcal{N}=1$ super Yang-Mills theory using the
  gradient flow},'' \href{http://dx.doi.org/10.1103/PhysRevD.100.074501}{{\em
  Phys. Rev. D} {\bfseries 100} no.~7, (2019) 074501},
  \href{http://arxiv.org/abs/1902.08469}{{\ttfamily arXiv:1902.08469
  [hep-lat]}}.

\bibitem{Hieda:2017sqq}
K.~Hieda, A.~Kasai, H.~Makino, and H.~Suzuki, ``{4D $\mathcal{N}=1$ SYM
  supercurrent in terms of the gradient flow},''
  \href{http://dx.doi.org/10.1093/ptep/ptx073}{{\em PTEP} {\bfseries 2017}
  no.~6, (2017) 063B03},
\href{http://arxiv.org/abs/1703.04802}{{\ttfamily arXiv:1703.04802 [hep-lat]}}.
%%CITATION = ARXIV:1703.04802;%%.

\bibitem{Kasai:2018koz}
A.~Kasai, O.~Morikawa, and H.~Suzuki, ``{Gradient flow representation of the
  four-dimensional $\mathcal{N}=2$ super Yang–Mills supercurrent},''
  \href{http://dx.doi.org/10.1093/ptep/pty117}{{\em PTEP} {\bfseries 2018}
  no.~11, (2018) 113B02},
\href{http://arxiv.org/abs/1808.07300}{{\ttfamily arXiv:1808.07300 [hep-lat]}}.
%%CITATION = ARXIV:1808.07300;%%.

\bibitem{Chigusa:2019wxb}
S.~Chigusa, T.~Moroi, and Y.~Shoji, ``{Bounce Configuration from Gradient
  Flow},'' \href{http://dx.doi.org/10.1016/j.physletb.2019.135115}{{\em Phys.
  Lett. B} {\bfseries 800} (2020) 135115},
  \href{http://arxiv.org/abs/1906.10829}{{\ttfamily arXiv:1906.10829
  [hep-ph]}}.

\bibitem{Sato:2019axv}
R.~Sato, ``{Simple Gradient Flow Equation for the Bounce Solution},''
  \href{http://dx.doi.org/10.1103/PhysRevD.101.016012}{{\em Phys. Rev. D}
  {\bfseries 101} no.~1, (2020) 016012},
  \href{http://arxiv.org/abs/1907.02417}{{\ttfamily arXiv:1907.02417
  [hep-ph]}}.

\bibitem{Hamada:2020rnp}
Y.~Hamada and K.~Kikuchi, ``{Obtaining the sphaleron field configurations with
  gradient flow},'' \href{http://dx.doi.org/10.1103/PhysRevD.101.096014}{{\em
  Phys. Rev. D} {\bfseries 101} no.~9, (2020) 096014},
  \href{http://arxiv.org/abs/2003.02070}{{\ttfamily arXiv:2003.02070
  [hep-th]}}.

\bibitem{Ho:2019ads}
D.~L.~J. Ho and A.~Rajantie, ``{Classical production of \textquoteright{}t
  Hooft\textendash{}Polyakov monopoles from magnetic fields},''
  \href{http://dx.doi.org/10.1103/PhysRevD.101.055003}{{\em Phys. Rev. D}
  {\bfseries 101} no.~5, (2020) 055003},
  \href{http://arxiv.org/abs/1911.06088}{{\ttfamily arXiv:1911.06088
  [hep-th]}}.

\bibitem{Fujikawa:2016qis}
K.~Fujikawa, ``{The gradient flow in $\lambda\phi^{4}$ theory},''
  \href{http://dx.doi.org/10.1007/JHEP03(2016)021}{{\em JHEP} {\bfseries 03}
  (2016) 021},
\href{http://arxiv.org/abs/1601.01578}{{\ttfamily arXiv:1601.01578 [hep-lat]}}.
%%CITATION = ARXIV:1601.01578;%%.

\bibitem{Morikawa:2018fek}
O.~Morikawa and H.~Suzuki, ``{Axial $U(1)$ anomaly in a gravitational field via
  the gradient flow},'' \href{http://dx.doi.org/10.1093/ptep/pty073}{{\em PTEP}
  {\bfseries 2018} no.~7, (2018) 073B02},
\href{http://arxiv.org/abs/1803.04132}{{\ttfamily arXiv:1803.04132 [hep-th]}}.
%%CITATION = ARXIV:1803.04132;%%.

\bibitem{Luscher:2013cpa}
M.~Luscher, ``{Chiral symmetry and the Yang--Mills gradient flow},''
  \href{http://dx.doi.org/10.1007/JHEP04(2013)123}{{\em JHEP} {\bfseries 04}
  (2013) 123},
\href{http://arxiv.org/abs/1302.5246}{{\ttfamily arXiv:1302.5246 [hep-lat]}}.
%%CITATION = ARXIV:1302.5246;%%.

\bibitem{Suzuki:2013gza}
H.~Suzuki, ``{Energy–momentum tensor from the Yang–Mills gradient flow},''
  \href{http://dx.doi.org/10.1093/ptep/ptt059, 10.1093/ptep/ptv094}{{\em PTEP}
  {\bfseries 2013} (2013) 083B03},
  \href{http://arxiv.org/abs/1304.0533}{{\ttfamily arXiv:1304.0533 [hep-lat]}}.
[Erratum: PTEP2015,079201(2015)].
%%CITATION = ARXIV:1304.0533;%%.

\bibitem{Makino:2014taa}
H.~Makino and H.~Suzuki, ``{Lattice energy–momentum tensor from the
  Yang–Mills gradient flow—inclusion of fermion fields},''
  \href{http://dx.doi.org/10.1093/ptep/ptu070, 10.1093/ptep/ptv095}{{\em PTEP}
  {\bfseries 2014} (2014) 063B02},
  \href{http://arxiv.org/abs/1403.4772}{{\ttfamily arXiv:1403.4772 [hep-lat]}}.
[Erratum: PTEP2015,079202(2015)].
%%CITATION = ARXIV:1403.4772;%%.

\bibitem{Asakawa:2013laa}
{\bfseries FlowQCD} Collaboration, M.~Asakawa, T.~Hatsuda, E.~Itou,
  M.~Kitazawa, and H.~Suzuki, ``{Thermodynamics of SU(3) gauge theory from
  gradient flow on the lattice},''
  \href{http://dx.doi.org/10.1103/PhysRevD.90.011501,
  10.1103/PhysRevD.92.059902}{{\em Phys. Rev.} {\bfseries D90} no.~1, (2014)
  011501}, \href{http://arxiv.org/abs/1312.7492}{{\ttfamily arXiv:1312.7492
  [hep-lat]}}.
[Erratum: Phys. Rev.D92,no.5,059902(2015)].
%%CITATION = ARXIV:1312.7492;%%.

\bibitem{Taniguchi:2016ofw}
Y.~Taniguchi, S.~Ejiri, R.~Iwami, K.~Kanaya, M.~Kitazawa, H.~Suzuki, T.~Umeda,
  and N.~Wakabayashi, ``{Exploring $N_{f}$ = 2+1 QCD thermodynamics from the
  gradient flow},'' \href{http://dx.doi.org/10.1103/PhysRevD.96.014509,
  10.1103/PhysRevD.99.059904}{{\em Phys. Rev.} {\bfseries D96} no.~1, (2017)
  014509}, \href{http://arxiv.org/abs/1609.01417}{{\ttfamily arXiv:1609.01417
  [hep-lat]}}.
[Erratum: Phys. Rev.D99,no.5,059904(2019)].
%%CITATION = ARXIV:1609.01417;%%.

\bibitem{Kitazawa:2017qab}
M.~Kitazawa, T.~Iritani, M.~Asakawa, and T.~Hatsuda, ``{Correlations of the
  energy-momentum tensor via gradient flow in SU(3) Yang-Mills theory at finite
  temperature},'' \href{http://dx.doi.org/10.1103/PhysRevD.96.111502}{{\em
  Phys. Rev.} {\bfseries D96} no.~11, (2017) 111502},
\href{http://arxiv.org/abs/1708.01415}{{\ttfamily arXiv:1708.01415 [hep-lat]}}.
%%CITATION = ARXIV:1708.01415;%%.

\bibitem{Yanagihara:2018qqg}
R.~Yanagihara, T.~Iritani, M.~Kitazawa, M.~Asakawa, and T.~Hatsuda,
  ``{Distribution of Stress Tensor around Static Quark--Anti-Quark from
  Yang-Mills Gradient Flow},''
  \href{http://dx.doi.org/10.1016/j.physletb.2018.09.067}{{\em Phys. Lett.}
  {\bfseries B789} (2019) 210--214},
\href{http://arxiv.org/abs/1803.05656}{{\ttfamily arXiv:1803.05656 [hep-lat]}}.
%%CITATION = ARXIV:1803.05656;%%.

\bibitem{Harlander:2018zpi}
R.~V. Harlander, Y.~Kluth, and F.~Lange, ``{The two-loop energy–momentum
  tensor within the gradient-flow formalism},''
  \href{http://dx.doi.org/10.1140/epjc/s10052-018-6415-7}{{\em Eur. Phys. J.}
  {\bfseries C78} no.~11, (2018) 944},
\href{http://arxiv.org/abs/1808.09837}{{\ttfamily arXiv:1808.09837 [hep-lat]}}.
%%CITATION = ARXIV:1808.09837;%%.

\bibitem{Iritani:2018idk}
T.~Iritani, M.~Kitazawa, H.~Suzuki, and H.~Takaura, ``{Thermodynamics in
  quenched QCD: energy–momentum tensor with two-loop order coefficients in
  the gradient-flow formalism},''
  \href{http://dx.doi.org/10.1093/ptep/ptz001}{{\em PTEP} {\bfseries 2019}
  no.~2, (2019) 023B02},
\href{http://arxiv.org/abs/1812.06444}{{\ttfamily arXiv:1812.06444 [hep-lat]}}.
%%CITATION = ARXIV:1812.06444;%%.

\bibitem{Capponi:2015ucc}
F.~Capponi, A.~Rago, L.~Del~Debbio, S.~Ehret, and R.~Pellegrini,
  ``{Renormalisation of the energy-momentum tensor in scalar field theory using
  the Wilson flow},'' \href{http://dx.doi.org/10.22323/1.251.0306}{{\em PoS}
  {\bfseries LATTICE2015} (2016) 306},
  \href{http://arxiv.org/abs/1512.02851}{{\ttfamily arXiv:1512.02851
  [hep-lat]}}.

\bibitem{Wess:1992cp}
J.~Wess and J.~Bagger, ``{Supersymmetry and Supergravity SECOND EDITION,
  REVISED AND EXPANDED},''
{\em Princeton Serieis in Physics} Princeton University Press, Princeton, New
  Jersey (1992).
%%CITATION = ARXIV:1302.5246;%%.

\bibitem{Wess:1973kz}
J.~Wess and B.~Zumino, ``{A Lagrangian Model Invariant Under Supergauge
  Transformations},''
  \href{http://dx.doi.org/10.1016/0370-2693(74)90578-4}{{\em Phys. Lett. B}
  {\bfseries 49} (1974) 52}.

\bibitem{Iliopoulos:1974zv}
J.~Iliopoulos and B.~Zumino, ``{Broken Supergauge Symmetry and
  Renormalization},''
  \href{http://dx.doi.org/10.1016/0550-3213(74)90388-5}{{\em Nucl. Phys. B}
  {\bfseries 76} (1974) 310}.

\bibitem{Grisaru:1979wc}
M.~T. Grisaru, W.~Siegel, and M.~Rocek, ``{Improved Methods for Supergraphs},''
  \href{http://dx.doi.org/10.1016/0550-3213(79)90344-4}{{\em Nucl. Phys. B}
  {\bfseries 159} (1979) 429}.

\bibitem{Seiberg:1993vc}
N.~Seiberg, ``{Naturalness versus supersymmetric nonrenormalization
  theorems},'' \href{http://dx.doi.org/10.1016/0370-2693(93)91541-T}{{\em Phys.
  Lett. B} {\bfseries 318} (1993) 469--475},
  \href{http://arxiv.org/abs/hep-ph/9309335}{{\ttfamily arXiv:hep-ph/9309335}}.

\end{thebibliography}

\providecommand{\href}[2]{#2}\begingroup\raggedright\endgroup

\end{document}